\documentclass[journal]{IEEEtran}

\usepackage{cite,amssymb}
\usepackage[nolist]{acronym}
\usepackage{graphicx}
\usepackage[cmex10]{amsmath}
\usepackage{subfigure}
\usepackage{color}
\usepackage{multirow}
\usepackage[normalem]{ulem}
\usepackage{soul}
\usepackage{hyperref}
\usepackage{mathtools}

\usepackage{amsmath}
\usepackage{algorithm}
\usepackage{algpseudocode}
\makeatletter
\def\BState{\State\hskip-\ALG@thistlm}
\makeatother

\newtheorem{lemma}{Lemma}
\newtheorem{proposition}{Proposition}

\def\Prob {{\mathbb P}}

\newcommand{\EXs}[2] {{\mathbb{E}}_{{#1}}\!\!\left\{{#2}\right\}}

\newcommand*{\density}{\rho}
\newcommand*{\densityRR}{\rho_\text{RR}}

\newcommand*{\CAMperiod}{T_\text{CAM}}
\newcommand*{\CAMfrequency}{f_\text{CAM}}
\newcommand*{\bBytes}{B_\text{CAM}}

\newcommand*{\nRes}{R}
\newcommand*{\nResf}{R_\text{f}}
\newcommand*{\nRest}{R_\text{t}}
\newcommand*{\res}[1]{r_{#1}}

\newcommand{\PTX}{P_{\text{t}}}
\newcommand{\PRX}{P_{\text{r}}}
\newcommand{\PRZERO}{P_{\text{r0}}}
\newcommand{\Gr}{G_{\text{r}}}
\newcommand{\Gt}{G_{\text{t}}}
\newcommand{\PN}{P_{\text{n}}}
\newcommand{\alphaPar}{{L}_\text{0}}
\newcommand{\betaPar}{{\beta}}
\newcommand{\chAtte}[1]{L(#1)}

\newcommand{\SNR}{\gamma}
\newcommand{\SNRmin}{{\gamma}_\text{m}}

\newcommand{\dist}{{d_\text{sd}}}
\newcommand{\distQoS}{{d_\text{0.9}}}
\newcommand{\distaverage}{{\bar d}}
\newcommand{\vardist}{\delta}

\newcommand{\varInterf}{y}

\newcommand{\prr}{{P_\text{RP}}}
\newcommand{\prrstar}{{P_\text{RP}^*}}

\newcommand{\subfuncMD}[2]{\xi\left(#1,#2\right)}
\newcommand{\subfuncMDder}[1]{\xi'\left(#1\right)}
\newcommand{\subfuncMDalpha}[2]{\xi_{\paramA,\paramB}\left(#1,#2\right)}
\newcommand{\subfuncMDalphader}[1]{{\xi_{\paramA}'\left(#1\right)}}
\newcommand{\subfuncMDalphaTriple}[3]{{\xi_{#1}\left(#2,#3\right)}}
\newcommand{\subfuncMDalphaderTriple}[2]{{\xi_{#1}'\left(#2\right)}}
\newcommand{\KRR}{K_\text{RR}}
\newcommand{\KMD}{K_\text{MD}}
\newcommand{\vardistn}{\Delta_n}

\newcommand{\fdistn}{f_{\vardistn}}
\newcommand{\Fdistn}{F_{\vardistn}}
\newcommand{\Fcompldistn}{\bar F_{\vardistn}}

\newcommand{\FdistR}{{F_{\Delta_R}}}
\newcommand{\Fdist}[1]{{F_{\Delta_{#1}}}}
\newcommand{\cFdistR}{{\bar F_{\Delta_R}}}

\newcommand{\PI}{I_\text{tot}}

\newcommand{\PIgenericTriple}[1]{I_{#1}}
\newcommand{\PIgeneric}{I_{{n,\paramA,\paramB}}}
\newcommand{\PIi}[1]{I_\text{#1}}
\newcommand{\PIleft}{I^\text{(l)}}

\newcommand{\PIright}{I^\text{(r)}}

\newcommand{\PImin}{I_\text{max}}
\newcommand{\distInterf}[1]{{{\vardist}_{\text{int}}^{(#1)}}}
\newcommand{\nInterf}{{n_\text{int}}}
\newcommand{\setInterferers}{\mathcal{N}_\text{int}}

\newcommand{\PhiY}{\Phi_{\PI}}
\newcommand{\PhiYa}{\Phi_{{\PI}_a}}

\newcommand{\PhiGa}{\Phi_{{\PIi{i}}_a}}

\newcommand{\G}[1]{\Upsilon(#1)}
\newcommand{\Ginv}[1]{{\Upsilon}^{-1}(#1)}
\newcommand{\GshiftedAlfa}[1]{  {{\Upsilon}^\text{(I)}_{\paramA,\paramB}}(#1)  }
\newcommand{\GshiftedAlfainv}[1]{{ {\Upsilon}^{\text{(I)}^{-1}}_{\paramA,\paramB}}(#1)}

\newcommand{\Pkeep}{p_\text{keep}}
\newcommand{\fading}{\upsilon}
\newcommand{\pHDloss}{P_\text{HD}}

\newcommand{\paramA}{\alpha_\text{1}}
\newcommand{\paramB}{\alpha_\text{2}}
\newcommand{\setValidy}{\mathcal{Y_\text{v}}}

\begin{acronym} 
\acro{5G}{fifth-generation} 
\acro{AWGN}{additive white Gaussian noise}
\acro{BEP}{beacon error probability}
\acro{BP}{beacon periodicity}
\acro{BSM}{basic safety message}
\acro{BSS}{basic service set}
\acro{C-ITS}{cooperative intelligent transport systems}
\acro{C-V2X}{cellular-\ac{V2X}}
\acro{CAM}{cooperative awareness message}
\acro{CAM-R}{\ac{CAM} resource}
\acro{CCA}{clear channel assessment}
\acro{CCDF}{complementary cumulative distribution function}
\acro{CD}{collision detection}
\acro{CDF}{cumulative distribution function}
\acro{c.f.}{characteristic function}
\acro{CSI}{channel state information}
\acro{CSMA/CA}{carrier sense multiple access with collision avoidance}
\acro{D2D}{device-to-device}
\acro{DSRC}{dedicated short range communication}  
\acro{ERP}{equivalent radiated power}
\acro{FCD}{floating car data}
\acro{FD}{full duplex}
\acro{FDD}{frequency division duplex}
\acro{GPS}{global positioning system}
\acro{HD}{half duplex}
\acro{i.i.d.}{independent identically distributed}
\acro{LOS}{line of sight}
\acro{LTE}{long term evolution}  
\acro{LTE-D2D}{\ac{LTE} with device-to-device communications}  
\acro{LTE-V2V}{\ac{LTE}-vehicle-to-vehicle}  
\acro{LTE-V2X}{\ac{LTE}-vehicle-to-anything}  
\acro{MAC}{medium access control}
\acro{MCS}{modulation and coding scheme}
\acro{MD}{allocation with maximum reuse distance}
\acro{NHTSA}{National Highway Traffic Safety Administration}
\acro{OBU}{on board unit}
\acro{OCB}{outside the context of a BSS}
\acro{OFDM}{orthogonal frequency division multiplexing}
\acro{PDF}{probability density function}
\acro{PEP}{pairwise error probability} 
\acro{PHY}{physical}
\acro{PMF}{probability mass function}
\acro{PPP}{Poisson point process}
\acro{ProSe}{proximity services}
\acro{PRP}{packet reception probability}
\acro{QoS}{quality of service}
\acro{r.v.}{random variable}
\acro{RF}{radio frequency} 
\acro{RB}{resource block}
\acro{RR}{random resource allocation}
\acro{RSU}{road side unit} 
\acro{SC-FDMA}{single carrier-frequency division multiple access}
\acro{SHINE}{simulation platform for heterogeneous interworking networks}
\acro{SI}{self-interference}
\acro{SINR}{signal to noise and interference ratio}
\acro{SNR}{signal to noise ratio}
\acro{SPS}{semi-persistent scheduling}
\acro{SV}{smart vehicle}
\acro{TDD}{time division duplex}
\acro{V2C}{vehicle-to-cellular} 
\acro{V2I}{vehicle-to-infrastructure} 
\acro{V2R}{vehicle-to-roadside}
\acro{V2V}{vehicle-to-vehicle} 
\acro{V2X}{vehicle-to-anything} 
\acro{VANETs}{vehicular ad hoc networks} 
\acro{WAVE}{wireless access in vehicular environment}
\end{acronym}

\begin{document}


%
\title{Analytical Investigation of Two Benchmark Resource Allocation Algorithms for LTE-V2V}
%
%
%



\author{$\mbox{Alessandro Bazzi}^{*}$,~\IEEEmembership{Senior Member,~IEEE,}
	Alberto Zanella,~\IEEEmembership{Senior Member,~IEEE,}\\
	Giammarco Cecchini,~\IEEEmembership{Student Member,~IEEE,}
     Barbara~M. Masini,~\IEEEmembership{Member,~IEEE}\\
     
\thanks{${}^{*}$Corresponding author.}
                \thanks{Alessandro Bazzi, Alberto Zanella, Giammarco Cecchini, and Barbara M. Masini
    are with
    CNR, IEIIT, Italy
    (e-mail: {\tt name.surname@cnr.it}).}
}

\maketitle

\begin{abstract}
Short-range wireless technologies will enable vehicles to communicate and coordinate their actions, thus improving people's safety and traffic efficiency. Whereas IEEE 802.11p (and related standards) had been the only practical solution for years, in 2016 a new option was introduced with Release 14 of long term evolution (LTE), which includes new features to enable direct vehicle-to-vehicle (V2V) communications. LTE-V2V promises a more efficient use of the channel compared to IEEE 802.11p thanks to an improved PHY layer and the use of orthogonal resources at the MAC layer. In LTE-V2V, a key role is played by the resource allocation algorithm and increasing efforts are being made to design new solutions to optimize the spatial reuse.In this context, an important aspect still little studied, is therefore that of identifying references that allow: 1) to have a perception of the space in which the resource allocation algorithms move; and 2) to verify the performance of new proposals. In this work, we focus on a highway scenario and identify two algorithms to be used as a minimum and maximum reference in terms of the packet reception probability (PRP). The PRP is derived as a function of various parameters that describe the scenario and settings, from the application to the physical layer. Results, obtained both in a simplified Poisson point process scenario and with realistic traffic traces, show that the PRP varies considerably with different algorithms and that there is room for the improvement of current solutions.
\end{abstract}



\IEEEpeerreviewmaketitle

\section{Introduction}\label{Section:intro}

Connected and automated vehicles promise to change people's lives over the next few years, with increased safety, more efficient traffic management and new services for drivers and passengers. As a complement to automation, the use of short-range wireless communications will improve awareness of the surrounding environment and allow cooperation between vehicles on the move and during maneuvers.
At the base of most applications, particularly in terms of safety, there are broadcast  messages, hereafter referred as \acp{CAM},\footnote{Such messages are called \acp{CAM} in European standardization \cite{3GPP_EN_302_637_2} and correspond to a subclass of the \acp{BSM} in the American specifications  \cite{SAE_DSRC_J2945_1}.} used by each vehicle to inform  neighbours of their position, direction, speed and so on.

Until late 2016, the main set of standards designed for short-range vehicular communications were the American \ac{WAVE} and the European counterpart \ac{C-ITS}/ITS-G5, both based on IEEE 802.11p at the \ac{PHY} and \ac{MAC} layers. At that time, 3GPP included in Release 14 of \ac{LTE} the so-called \ac{LTE-V2X}\acused{V2X} or, more generally, \ac{C-V2X} to allow direct short-range communications between vehicles and other devices on the road. In particular, the subset represented by the communication between vehicles takes the name of \ac{LTE-V2V}\acused{V2V}. LTE-V2V promises broader coverage and more efficient use of wireless resources than IEEE 802.11p  \cite{SeoLeeYasPenSar:J16,SunHuPenPanZhaFan:J16,BazMasZanThi:J17}. The longer range is possible thanks to the improved channel coding and the possible partial use of the bandwidth (less noise at the receiver), while the greater efficiency is a consequence of the \ac{MAC} layer characterized by orthogonal resources.

More specifically, LTE-V2V is based on an organization of time and frequency domains in orthogonal resources; nodes that use different combinations cause negligible reciprocal interference. One of the main problems is therefore the design of an efficient resource allocation scheme that guarantees a different assignment to the nodes located close to each other.  Clearly, the specific algorithm plays a key role in system performance.

Very recently, a number of works concerning the resource allocation algorithms for LTE-V2V have been published both by assuming network support \cite{ZhaHouXuTao:C16,AbaKopHee:C17,CecBazMasZan:C17_2,HuEicDilBotGoz:C16} and a distributed approach \cite{MolGoz:J17,YanPelCha:C16,KimLeeMooHwa:J18,CecBazMasZan:C17_3,HeTangFanZhang:L18}. Obviously, every time a new algorithm has been proposed, an improvement has been shown with respect to some reference; however: 1) there is no agreement on the references to be used; and 2) a discussion about the distance from an optimal allocation is not present. Furthermore, the \ac{PRP} is normally estimated by means of simulations and no analytical expressions are provided. 

To deal with this gap, in this paper we  focus on CAM transmissions in a highway scenario and identify two algorithms to be used as references. For both of them, we derive the \ac{PRP} analytically. The first reference 
corresponds to a random allocation, in which each assignment is performed without any knowledge of the resources occupied by the other nodes; this algorithm represents a pessimistic solution, to be used as an inferior benchmark. 
The second exploits the exact position of all the vehicles and assigns resources in order, thus maximizing the average distance between the interfering nodes (as better discussed later); this algorithm represents an optimistic solution, to be used as a superior benchmark.

The contribution of the paper is based on the following five steps.
\begin{enumerate}
	\item We formalize the \ac{PRP} evaluation in LTE-V2V in a highway scenario, formulating the problem based on the source-destination distance;
	\item We focus on a random assignment as a pessimistic reference, indicated in the further as \textit{basic reference}, and derive the relative \ac{PRP}; 
	\item We indicate the scheme that maximizes the average distance between the interfering nodes, allocating the resources following the order of the node positions, as an optimistic benchmark, denoted as \textit{maximum reuse reference}, and derive the relative \ac{PRP};
	\item We provide examples of results to demonstrate the correctness of the analysis for two different highway scenarios: in the first one, the positions of the vehicles are modelled as a homogeneous 1-D \ac{PPP}; in the second one, realistic traffic traces are adopted;
	\item We also show example results that compare the two benchmarks with algorithms based on specifications and literature.
\end{enumerate}

\begin{table*}[t]
\caption{Related work dealing with resource allocation in LTE-V2V.
\label{Tab:RelatedWork}}
\scriptsize
\centering
\begin{tabular}{p{2.8cm}|p{1.9cm}p{4.0cm}p{2.3cm}p{4.7cm}}
\hline
\textbf{Reference} & \textbf{LTE Release} & \textbf{Algorithm proposed} & \textbf{Evaluation Tool} & \textbf{Benchmark(s)} \\ \hline 
Zhang et al. \cite{ZhaHouXuTao:C16} & Before Release 14 & Controlled algorithm, based on CSI & Not specified simulator & i) A modified algorithm from the literature and ii) random allocation with check of the SINR\\
Abanto-Leon et al. \cite{AbaKopHee:C17} & Release 14 & Controlled algorithm based on clustering and graph-theory & Not specified simulator & Exhaustive search\\
Cecchini et al. \cite{CecBazMasZan:C17_2} & Release 14 & Controlled algorithm, based on reuse distance & LTEV2Vsim \cite{CecBazMasZan:C17} & No benchmark \\
Hu et al. \cite{HuEicDilBotGoz:C16} & Before release 14 & Controlled algorithm, based on graph coloring & Not specified simulator & An ideal scheme, without interference \\
Molina-Menegosa et al. \cite{MolGoz:J17} & Beyond Release 14 & Autonomous, modified from 3GPP & Veins \cite{SomGerDre:J11} & Standard 3GPP autonomous algorithm \\
Yang et al. \cite{YanPelCha:C16} & Before Release 14 & Autonomous, based on positions & Not specified simulator & Random allocation \\ 
Kim et al. \cite{KimLeeMooHwa:J18} & Before Release 14 & Autonomous, based on positions & Not specified simulator & Random allocation \\ 
Cecchini et al. \cite{CecBazMasZan:C17_3} & Beyond Release 14 & Autonomous,  with observed occupations piggybacked in data packets & LTEV2Vsim \cite{CecBazMasZan:C17} & Two algorithms from the literature \\
He et al. \cite{HeTangFanZhang:L18} & Beyond Release 14 & Autonomous, with reservations piggybacked in data packets  & Ad-hoc simulator & Random allocation and basic enhancements \\
\hline
\end{tabular}
\end{table*}


The rest of the paper is organized as follows. Section~\ref{Section:relatedwork} gives a brief overview of LTE-V2V and related work. In Section~\ref{Section:systemmodel}, the notation, the scenario and the assumptions are detailed and the problem is formally defined. In Section~\ref{Section:prrcalculation}, the two reference algorithms are discussed in detail and the \ac{PRP} is derived. The numerical results that demonstrate the validity of the analysis and show some relevant examples are provided in Section~\ref{Section:results}. Finally, our conclusions and a discussion of directions for future improvements are provided in Section~\ref{Section:conclusion}.

\section{LTE-V2V and Related Work}\label{Section:relatedwork}

\subsubsection{LTE-V2V physical and MAC layers} The first C-V2X specifications are included in Release~14, frozen in June 2017 \cite{SeoLeeYasPenSar:J16,SunHuPenPanZhaFan:J16}. They are based on \ac{D2D} communication, defined as part of the \ac{ProSe} since Release~12. Direct \ac{V2V} communications, also referred to as \textit{sidelink}, use \ac{SC-FDMA} at the PHY and MAC layers (same as the \ac{LTE} uplink). In the frequency domain, the available bandwidth is separated into groups of orthogonal sub-carriers (12~contiguous sub-carriers, spaced 15~kHz apart from each other). In the time domain, the signal is separated into 10~ms frames, which are in turn subdivided into 10 subframes of 1~ms. A subframe normally accommodates hosts 14 \ac{OFDM} symbols, of which nine contain data, four are used for channel estimation and synchronization, and the last one is not used to allow timing adjustments and Tx-Rx switching.

The minimum unit for the allocation of resources is the pair of \acp{RB}, corresponding to a group of sub-carriers in the frequency domain and a subframe in the time domain. Depending on the adopted \ac{MCS}, the \ac{RB} pair carries a variable number of data bits, as described for example in \cite{BazMasZan:C17}; it follows that the number of \acp{RB} needed to allocate a message depends both on the size of the packet and on the MCS adopted.  

In LTE-V2X, sidelink resource allocation can be performed using one of the following two approaches, defined Mode~3 and Mode~4. 
In Mode~3, the allocation is performed by a central resource management entity that communicates decisions to individual devices. This mode is only possible if the devices are under coverage of the cellular network. In Mode~4, the allocation is performed autonomously by each vehicle and is therefore completely distributed.

Since the main service is the transmission of \acp{CAM}, which is by nature periodic with messages of constant size, a mechanism called \ac{SPS} is used; the \ac{SPS} implies that the same resources are allocated periodically for a certain time, minimizing the load caused by the signaling. While a specific algorithm is defined in Release~14 for Mode~4 \cite{3GPP_TS_36_213,3GPP_TS_36_321}, the algorithm to be used in Mode~3 is left to the network operator. 

\subsubsection{Related work} Several papers have recently addressed the comparison between LTE-V2V and IEEE~802.11p. When and how much LTE-V2V outperforms IEEE~802.11p, it is still under discussion. While some works, such as \cite{MinWinZhaBlaEtc:C17,NguShaSudKapEtc:C17}, show significant improvements with the new technology, others observe that it also depends on the specific scenario and settings. For example, both \cite{BazMasZanThi:J17} and \cite{MolGoz:J17} show that under high load conditions (heavy traffic and/or frequent messages), specific IEEE 802.11p settings may result in a higher \ac{PRP}. 

The impact of the resource allocation algorithm on system performance is indeed so relevant that several studies have recently focused on it, even in the absence of a comparison with IEEE~802.11p. While the first works (such as \cite{CheYanShe:J15,SunStrBraSouSui:J16,HunZhaFesCheFet:C16}) were mainly focused on the underlay scenario, where the same resources can be shared by both V2I and V2V connections, it is now assumed that specific resources will be reserved for V2V communications and attention has been moved more to the spatial reuse of resources. 

Examples focused on controlled allocations (Mode~3) are presented in \cite{ZhaHouXuTao:C16,AbaKopHee:C17,CecBazMasZan:C17_2,HuEicDilBotGoz:C16}. In \cite{ZhaHouXuTao:C16}, the authors propose an optimized allocation method based on the continuous transmission of the \ac{CSI} (criticized by some works as not feasible in vehicular scenarios \cite{CheYanShe:J15}). In \cite{AbaKopHee:C17}, the vehicles are clustered and then algorithms based on graph theory are applied to minimize allocation collisions. In \cite{CecBazMasZan:C17_2}, the allocation is performed based on a pre-defined reuse distance and vehicle location. In \cite{HuEicDilBotGoz:C16}, again taking advantage of the position of the nodes, a graph colouring methodology is adopted to correctly select the resource that each vehicle should use based on the assignment of its neighbours.

Autonomous allocations (Mode~4) are investigated in \cite{MolGoz:J17,YanPelCha:C16,KimLeeMooHwa:J18,CecBazMasZan:C17_3,HeTangFanZhang:L18}. While the algorithm specified by 3GPP \cite{3GPP_TS_36_213,3GPP_TS_36_321} is addressed in \cite{MolGoz:J17}, the others focus on new approaches. In both \cite{YanPelCha:C16} and \cite{KimLeeMooHwa:J18}, resources are grouped into sub-pools that are associated with vehicles based on their direction and/or position on the road, thus reducing the risk of collisions due to allocations performed by terminals hidden to each other (mainly because of the distance in the highway scenarios and of the buildings in the urban areas). In \cite{CecBazMasZan:C17_3}, the adoption of maps of the resources sent within the \acp{CAM} is proposed. In  \cite{HeTangFanZhang:L18}, the performance of autonomous selection algorithms is improved by piggybacking information about reservations with the data packets.

The main aspects of the cited papers dealing with resource allocations are also summarized in Table \ref{Tab:RelatedWork}. In all cases, the results are compared with simpler solutions and it is not possible to understand how much the obtained  performance is far from that of an optimal allocation. Furthermore, the results are always obtained by simulation. 

Differently, in this work we derive the \ac{PRP} for a basic and a maximum reuse allocation, with the aim to provide reference indications for the evaluation of new algorithms.

\section{System Model}\label{Section:systemmodel}

In this Section, the notation, the scenario, and the assumptions adopted are first detailed, followed by the formulation of the problem.

\subsection{Notation}\label{Subsection:notations}
Throughout the paper, $\Prob\{{\cal A}\}$ and $\mu_X\triangleq \EXs{X}{\cdot}$ indicate the
probability of the event $\cal A$, and the expectation with respect to the \ac{r.v.} $X$, respectively. 
 The functions $f_X(x)$, $F_X(x)$, and $\bar F_X(x)$ respectively indicate the \ac{PDF}, the \ac{CDF} and the \ac{CCDF} ($\Prob\{X > x\}$) of the \ac{r.v.} $X$. Finally, $g'(x) \triangleq dg(x)/dx$ is the derivative of the function $g$ with respect to the variable $x$. 

\subsection{Scenario and assumptions}\label{Subsection:scenario}

\subsubsection{Scenario} 
\begin{figure} [t]
	\centering
	\subfigure[Example 2-D scenario.]{
		\includegraphics[trim= {0 650 0 0}, clip,width=0.95\linewidth,draft=false]{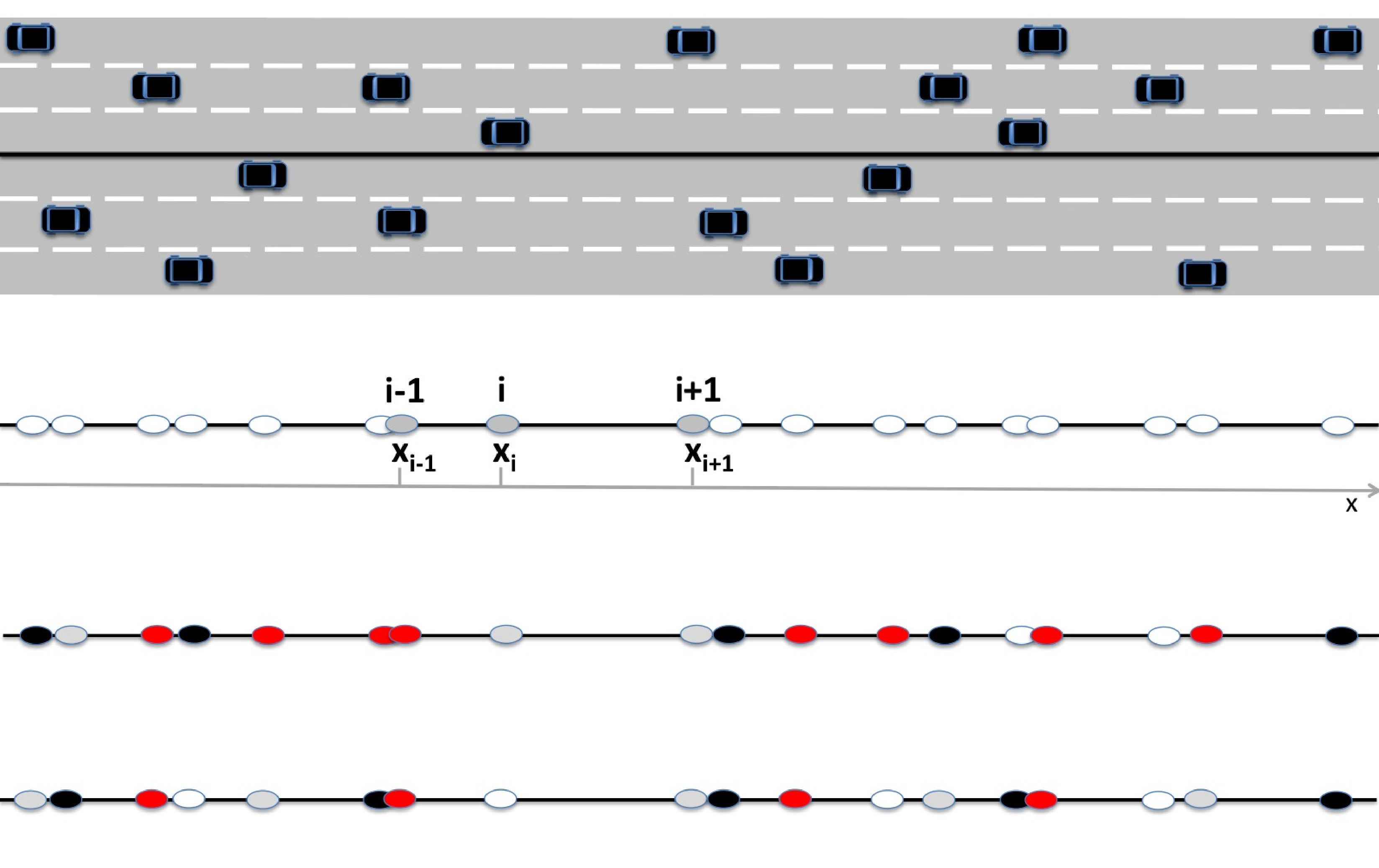}\label{fig:scenario_a}}
	\subfigure[1-D representation.]{
		\includegraphics[trim= {0 380 0 400}, clip,width=0.95\linewidth,draft=false]{Figures/Diapositiva1.eps}\label{fig:scenario_b}}
	\subfigure[Example of random resource allocation (RR).]{
		\includegraphics[trim= {0 200 0 700}, clip,width=0.95\linewidth,draft=false]{Figures/Diapositiva1.eps}\label{fig:scenario_c}}
	\subfigure[Example of allocation with maximum reuse distance (MD).]{
		\includegraphics[trim= {0 0 0 900}, clip,width=0.95\linewidth,draft=false]{Figures/Diapositiva1.eps}\label{fig:scenario_d}}
	\caption{Example 2-D scenario, 1-D representation, and example allocations. Allocations are shown assuming $\nRes=4$, with different colors representing orthogonal resources.}
	\label{fig:scenario}
\end{figure}

In this work, we assume a highway segment with variable traffic conditions, approximated as a \mbox{1-D} scenario with vehicles distributed according to Poisson,  
i.e., the vehicle positions follow a homogeneous \ac{PPP}. It has been shown that this approximation is in good agreement with the real distributions (refer, for example to \cite{CheRefMa:C07,ZhaCheYanEtAl:J12}). 
The validity of our results is anyway discussed in Section~\ref{Section:results} also with reference to a realistic highway scenario. In the following, $x_n$ indicates the coordinate of vehicle $n$; without loss of generality, the vehicles in the scenario are numbered in such a way that $x_i<x_j$ if $i<j$. An example of a 2-D scenario with its 1-D representation is provided in Figs.~\ref{fig:scenario_a} and~\ref{fig:scenario_b}.

\subsubsection{Resources} Each vehicle is equipped with an \ac{OBU} that periodically generates \acp{CAM} at a certain frequency $\CAMfrequency$,  i.e., with a given repetition interval  $\CAMperiod=1/\CAMfrequency$. \acp{CAM} are all of $\bBytes$ bytes. The allocation of a CAM requires the reservation of a group of contiguous RBs (as calculated for example in \cite{BazMasZan:C17}). Given the periodic nature of packet generation and the fixed size of the messages, during each $\CAMperiod$ it is possible to assign a fixed maximum number of CAMs using a deterministic number of groups of RBs. Such groups are hereafter denoted as \acp{CAM-R} and the number of \acp{CAM-R} per $\CAMperiod$ is denoted as $\nRes$. Each message is transmitted in one of the $\nRes$ available \acp{CAM-R} and any two devices interfere with each other if and only if they use the same \ac{CAM-R}.

Given the use of SC-OFDMA, $\nRes$ corresponds to a grid of $\nResf$ resources in the frequency domain and $\nRest$ resources in the time domain (i.e., $\nRes=\nResf\cdot\nRest$). As for time and frequency domains, resources are ordered so that the generic resource $r^\text{*}\in \{1,\nRes\}$ occupies in the time domain the slot $r^\text{*}_\text{t} = r^\text{*} \mod \nRest$ and in the frequency domain the portion $r^\text{*}_\text{f} = \lceil \frac{r^\text{*}}{\nRest} \rceil$ (i.e., $r^\text{*}=\left(r^\text{*}_\text{f}-1\right)\cdot\nRest + r^\text{*}_\text{t}$). This means, in particular, that two generic resources $r_a$ and $r_b$ are transmitted in the same time interval if $|r_a-r_b|$ is a multiple of $\nRest$.

\subsubsection{Propagation} The signal attenuation due to propagation is modelled as
\begin{equation}\label{eq:pathloss}
\chAtte{\vardist} = \frac{ \alphaPar{} \cdot \vardist^{\betaPar{}} }{ \fading }
\end{equation}
where $\alphaPar$ is the average path loss at the reference distance of 1\;m, $\vardist$ is the distance between the transmitter and the receiver, $\betaPar$ is the path loss exponent, and $\fading$ is a random variable that takes into account the channel variability, assumed constant during the transmission of a CAM.

Consider a generic transmitter, a generic receiver at distance $\vardist$, and the set of all interferers $\setInterferers$ (depending on the allocation algorithm). Each interferer $i \in \setInterferers$ is at a distance $\distInterf{i}$ from the receiver and with an independent channel variability represented by $\fading^{(i)}$. All signals are transmitted with the same power $\PTX$.  The \ac{SINR} at the receiver is calculated as 
\begin{equation}\label{eq:SINR}
\SNR = \frac{\PRX}{\PN + \PI} = \frac{ \frac{\PTX\cdot \Gt \cdot \Gr \cdot \fading}{\alphaPar \cdot \vardist^{\betaPar}} } { \PN + \sum\limits_{i \in \setInterferers}{\frac{\PTX\cdot \Gt \cdot \Gr \cdot \fading^{(i)}}{\alphaPar \cdot \distInterf{i}^{\betaPar{}}}}}
\end{equation}
where $\Gt$ is the antenna gain at the transmitter, $\Gr$ is the antenna gain at the receiver, $\PRX=\frac{\PTX \cdot \Gt \cdot \Gr \cdot \fading}{\alphaPar \cdot {\vardist}^{\betaPar}}$ is the power received from the desired transmitter, $\PN$ is the noise power, and $\PI=\sum\limits_{i \in \setInterferers}{\frac{\PTX \cdot \Gt \cdot \Gr\cdot \fading^{(i)}}{\alphaPar \cdot \distInterf{i}^{\betaPar{}}}}$ is the total interference power.

It is assumed that a message is correctly decoded if $\SNR$ is greater than a minimum threshold $\SNRmin$ (as for example in \cite{HunZhaFesCheFet:C16,ZhaCheYanEtAl:J12,CheRefMa:C07,HasVuSak:J11,MarSorAguKovGom:J18}).

\subsubsection{Half duplexing} Since the radios of the LTE-V2V devices are \ac{HD}, a node is not able to decode the message in the same subframe that it is using for transmissions. To take this into account, we define the probability of loss due to \ac{HD} as $\pHDloss$.

\subsection{Problem formulation}\label{Subsection:problemformulation}

Throughout this paper, the following definitions apply.
\begin{itemize}
	\item \textit{Source-to-destination distance}, $\dist$: it is the physical distance between the generic source and the generic destination;
	\item \textit{\ac{PRP}}, $\prr$: it is the probability that a vehicle at a given distance from the source correctly receives and decodes a message; based on the previous definition of correct decoding and taking into account the \ac{HD} problem, it is \begin{align}
	\prr=\left(1-\pHDloss\right) \cdot \Prob\{\SNR > \SNRmin\}\;.
	\end{align}
\end{itemize}

The purpose of this work is to derive $\prr$ as a function of~$\dist$, with the specified settings. All CAMs are considered of the same importance for all receivers, which is a common assumption. In any case, providing results based on the distance between source and destination allows us to differentiate the performance based on how close the vehicles are communicating. It is intuitive, in fact, that the relevance of the information is different if vehicles are close or far from each other; this consideration is also reflected in the requirements for V2V applications, which often include a maximum communication distance \cite{ETSI_102_638,MasBazZan:J18}.

\section{Benchmark Resource Allocation Algorithms and Packet Reception Probability Evaluation}\label{Section:prrcalculation}

In this Section, the two reference allocations are detailed and the corresponding \ac{PRP} is calculated, preceded by a discussion on the approach adopted and on the common expressions.

\subsection{Common calculations}\label{Subsection:preliminarycalculation}

The model expressed by \eqref{eq:SINR} includes independent random variables to describe the channel of the useful signal and that of each interfering signal; this makes the conventional mathematical methods not applicable to obtain the desired distribution. To simplify the problem, let us assume that the variability of the channel of the interferers is negligible, i.e. we assume $\fading^{i}=1,\; \forall i \in [1,\nInterf]$. Please note that this approximation is applied only to the analysis and not to the simulations used to validate the model. We also define the received power without the channel variability as $\widetilde{\PRX} \triangleq \frac{\PTX \cdot \Gt \cdot \Gr}{\alphaPar \cdot {\vardist}^{\betaPar}}$ and the maximum acceptable interference level for a given $\fading$ as
\begin{align}\label{eq:PImin}
\PImin(\fading) \triangleq \frac{\PRX(\fading)}{\SNRmin} - \PN = \frac{\widetilde{\PRX} \cdot \fading}{\SNRmin} - \PN\;.
\end{align}
Moreover, we denote as $\prrstar(\fading)$ the probability of correctly decoding a packet, ignoring the problem of the \ac{HD} and given a specific value of $\fading$.
According to \eqref{eq:SINR}, \eqref{eq:PImin}, and the definition of \ac{PRP}, it follows
\begin{align}\label{eq_Prrstar}
\prrstar(\fading)&= \Prob\left\{\PI  < \frac{\PRX(\fading)}{\SNRmin} - \PN \right\}\nonumber \\ &=\Prob \{ \PI < \PImin(\fading) \} = F_{\PI}(\PImin(\fading))\;.
\end{align}
Finally, since losses due to half duplexing and incorrect decoding are independent of each other, it is
\begin{align}\label{eq_Prr}
\prr&=\left(1-\pHDloss\right) \cdot \int_{-\infty}^{\infty}\prrstar(\fading)\Prob(\fading) d \fading\nonumber\\
&=\left(1-\pHDloss\right) \cdot  \int_{-\infty}^{\infty}F_{\PI}(\PImin(\fading))\Prob(\fading) d \fading \nonumber\\
&=\left(1-\pHDloss\right) \cdot  \int_{-\infty}^{\infty}F_{\PI}(\frac{\widetilde{\PRX} \cdot \fading}{\SNRmin} - \PN)\Prob(\fading) d \fading.
\end{align}

As a consequence, for any given distribution of the channel variability $\Prob(\fading)$, the solution to the problem defined in Section~\ref{Subsection:problemformulation} is obtained once the value of $\pHDloss$ and the expressions of  $F_{\PI}(y)$ are derived. Being $y$ interference, it is $y \geq 0$; this assumption is left implicit in the following.

\subsection{Random resource allocation (RR)}\label{Subsection:randomcalculation}

\textbf{\textit{Algorithm and motivation.}} We denote as \ac{RR} the selection of resources carried out randomly with uniform distribution. Formally, the \ac{CAM-R} allocated to node $n$ is
\begin{equation}
r_n=\text{randint}\{1,\nRes\}
\end{equation} 
where $\text{randint}\{a,b\}$ corresponds to a uniformly random selection of an integer between $a$ and $b$. A new selection is performed in each beacon interval. An example of RR is shown in Fig.~\ref{fig:scenario_c}, with $\nRes=4$. Among all the possible allocations, \ac{RR} represents a useful reference, since it does not exploit any parameter or knowledge, except for $\nRes$. 
If another algorithm performs worse or even slightly better, it is more convenient to replace it with \ac{RR} to reduce memory occupancy and computational load.

\textbf{\textit{Maximum interference calculation.}} Since the nodes are positioned according to a 1-D homogeneous \ac{PPP} and the allocation of resources is completely random, we can apply the Marking Theorem \cite{King:93} to demonstrate that the set of nodes that adopt a specific resource still follows a 1-D homogeneous \ac{PPP}. As a consequence, the interfering nodes still follow a \mbox{1-D PPP} with density $\densityRR=\density/\nRes$ and the interference can be calculated using the same approach as \cite{SalZanJ09,HaeGan:J09}, leading to the following Proposition.

\begin{proposition}
	The total interference with RR corresponds to a stable distribution	with parameters $\mu=0$, $a=1/\betaPar$, $b=1$, and $c=\PRZERO \left(2\densityRR \Gamma\left(\frac{\betaPar-1}{\betaPar}\right) \cos\left(\frac{\pi}{2 \betaPar}\right) \right)^\betaPar$, where $\PRZERO$ is defined in \eqref{eq:Pr0} and $\Gamma\left(x\right)=\int_{0}^{\infty} t^{x-1} e^{-t} dt$ is the gamma function. 
\end{proposition}

\begin{IEEEproof}
	The Proof is provided in Appendix A.
\end{IEEEproof}

Although there is no closed form (with the exception of case $\betaPar=2$, detailed in the Lemma that follows), the \ac{CDF} of the distribution can easily be obtained numerically, for example as proposed in \cite{Nol:J97}.

\begin{lemma}
If $\betaPar=2$, the \ac{CDF} can be written as 
\begin{align}\label{eq:Flemma}
F_{\PI}(y) = \text{erfc}\left( \densityRR \Gamma\left(1/2\right)\sqrt{\frac{\PRZERO}{y}}\right) \;.
\end{align}
\end{lemma}

\begin{IEEEproof}
	The Proof is provided in Appendix B.
\end{IEEEproof}

\textbf{\textit{Probability of the HD issue.}} Since the selection of the resource is random, the probability that the source and the destination use a resource of the grid that occupies the same time unit is simply
\begin{align}
\pHDloss = 1/\nRest\;.
\end{align}

\subsection{Allocation with maximum reuse distance (MD)}\label{Subsection:optimalcalculation}

\textbf{\textit{Algorithm and motivation.}} We denote as \ac{MD} the allocation obtained by ordering the vehicles based on their position and allocating the resources in cyclic order. Formally, the \ac{CAM-R} allocated to node $n$ is
\begin{equation}
r_n=n\mod\nRes
\end{equation} 
where $a\mod b$ is used to indicate $a$ modulo $b$ (i.e., the rest of the Euclidean division of $a$ by $b$). The sorting and allocation process is repeated at each beacon interval.
An example of \ac{MD} is shown in Fig.~\ref{fig:scenario_d}, with $\nRes=4$.  
\ac{MD}, inspired by the frequency planning in cellular networks, represents a relevant reference, as it maximizes the average distance between nodes using the same resource, as clarified in the following Proposition. 

\begin{proposition}\label{Theorem:optimality} In a 1-D PPP, given the number of resources $\nRes$, the maximum value of the average distance between nodes using the same resource is obtained by sorting the nodes based on their position and then allocating the resources in order, modulo $\nRes$. 
\end{proposition}

\begin{IEEEproof}
	The Proof is provided in Appendix C.
\end{IEEEproof}

Please note that, if the channel variability is neglected and taking into account that the interference is monotonically decreasing with distance, the same reasoning can be applied to show that the average interference mutually caused by the nodes is minimized. It should also be observed that the MD algorithm is proposed only as a reference, since its implementation is hardly realistic; in fact, it would require real-time knowledge of the position of all nodes by an entity (e.g., the network) and a continuous communication of the new allocation from the same entity to all nodes.

\textbf{\textit{Maximum interference calculation.}} With MD the set of interfering nodes is no longer a PPP. In this case, the assessment of the contribution of all the interferers does not appear to be tractable. To obtain a closed-form expression, we approximate the distribution of the interference at the generic destination by considering only the first interferer on the right and the first interferer on the left of the destination node. All other interferers are further away from the destination and are considered negligible. With this approximation, the following Proposition is valid.

\begin{proposition}
The \ac{CDF} of the received interference with MD can be approximated by\footnote{Equation \eqref{eq:cdfIMD} is derived with the following limitations, which are normally acceptable: 1) the probability that there is an interferer between source and destination is negligible; and 2) the useful received power is greater than the minimum acceptable interference. The general solution is also provided in Appendix D.}
\begin{align}\label{eq:cdfIMD}
F_{\PI}(\varInterf) &\approx \KMD \int_{0}^\varInterf \Gamma\left(\nRes,\density \left(\subfuncMD{\varInterf-z}{-\dist}\right)\right) \nonumber\\& \;\;\;\; \cdot
\left(\subfuncMD{z}{\dist}\right)^{\nRes-1} \subfuncMDder{z} e^{-\density \left(\subfuncMD{z}{\dist}\right)} dz
\end{align}	
where
$\KMD = -\frac{\density}{\left(\Gamma(\nRes)\right)^2}$, $\Gamma(s,x) \triangleq \int_{x}^{\infty} t^{s-1} e^{-t} d t$ is the upper incomplete gamma function, $\subfuncMD{a}{b} = \left(\frac{a}{\PRZERO}\right)^{-\frac{1}{\betaPar}}+b$, and $\subfuncMDder{a} = \frac{d\subfuncMD{a}{b}}{d a}=\left(-\frac{1}{\betaPar}\right)\PRZERO^{1/\betaPar}a^{-\frac{1+\betaPar}{\betaPar}}$.
\end{proposition}

\begin{IEEEproof}
	The proof is provided in Appendix D.
\end{IEEEproof}

\textbf{\textit{Probability of the HD issue.}} Since the nodes are sorted and the resources are ordered as specified in Section~\ref{Subsection:scenario}, the probability that the receiver transmits in the same time interval as the source corresponds to the probability that the number of nodes between them is equal to $\left(\nRest-1\right) + k \cdot \nRest$, for any integer $k \geq 0$. Given the hypothesis of 1-D PPP, this probability can be derived as
\begin{align}
\pHDloss &= \sum_{k=0}^{\infty}\Prob\{\text{to have}~\left(\nRest-1\right) + k \cdot \nRest~\text{nodes} \in \left[0,\dist\right]\} 
\nonumber \\&
=\sum_{k=0}^{\infty} \frac{\left(\density\dist\right)^{\left(\nRest-1\right) + k \cdot \nRest}}{\left((\nRest-1) + k \cdot \nRest\right)!} e^{-\density \dist}\;.
\end{align}
 
\section{Numerical Results}\label{Section:results}

In this Section, example results are shown to validate the models and demonstrate the efficacy of the proposals as references, first in a 1-D PPP scenario and then considering a realistic highway traffic trace.

\begin{table}
\caption{Main settings.}
\vspace{-2mm}
\label{Tab:settings}
\centering
\begin{tabular}{p{5.5cm}|p{2.5cm}}\hline
\textbf{Parameter (Symbol)} & \textbf{Value} \\  \hline
Vehicle density  ($\rho$) &  0.1 vehicles/m (*) \\
Source to destination distance ($\dist$) & 120\;m (*) \\ 
Beacon frequency & 10 Hz (*) \\
Transmitted power ($\PTX$) & 23 dBm \\
Antenna gain at the transmitter ($\Gt{}$) & 3 dB \\
Antenna gain at the receiver ($\Gr{}$) & 3 dB \\
Path loss at 1 m ($\alphaPar{}$) & 20.06 dB \\
Loss exponent ($\betaPar{}$) & 4 \\
Noise figure & 9 dB \\
Bandwidth & 10 MHz \\
Pairs of RBs per subframe for data & 40 \\
MCS & 4 \\ 
Modulation & QPSK \\ 
Beacon size ($\bBytes$) & 300 bytes\\ 
RBs per beacon & 68 \\
Beacon resources per beacon period ($\nRes$) & 100 (*) \\
Minimum SINR ($\SNRmin$) & 2.8 dB \\
\hline
\multicolumn{2}{l}{(*) Value used when not differently specified} \\
\end{tabular}
\end{table}

\subsection{Main settings}

The main settings are summarized in Table~\ref{Tab:settings}. In accordance with 3GPP in \cite{3GPP_TR_36_885}, we have adopted $\bBytes=300$~bytes, a bandwidth of 10~MHz, $\Gt=\Gr=3$~dB, a noise figure of 9~dB, and the WINNER+ B1 model \cite{WINNERplus} for propagation, corresponding to $\alphaPar{}=20.06$~dB and $\betaPar=4$.\footnote{In principle, the WINNER+ B1 model \cite{WINNERplus} has a dual-slope, with the given values that are valid for distances larger than about 20~m. However, this distance is very small in the considered scenario. For simplicity, the analysis is therefore approximated as single-slope. In any case, all the simulations were performed using the dual-slope model.} To model the channel variability, we assume the presence of log-normal shadowing with a standard deviation of 3~dB, as recommended by 3GPP in \cite{3GPP_TR_36_885}; in the simulations, the shadowing is correlated, with decorrelation distance 25\;m. A transmission power of $\PTX=23$~dBm is assumed, equal to the maximum transmission power allowed in \cite{3GPP_TS_36_101}.

MCS~4 is adopted with the following rationale. Assuming as in \cite{BazMasZanThi:J17} that 40 \ac{RB} pairs are available in each LTE subframe for CAM allocation, MCS~4 is the most reliable MCS that does not require more than one subframe to allocate a beacon of $\bBytes=300$~bytes (each CAM uses 68 RBs). The allocation of a CAM per subframe ($\nResf=1$) leads to a number of \acp{CAM-R} $\nRes=\nRest$ and a threshold $\SNRmin=2.8\;dB$ (obtained as in \cite{BazMasZan:C17}). Unless otherwise specified, a beacon frequency of 10\;Hz is assumed, which is the most commonly adopted value and corresponds to $\nRes=\nRest=100$ (recalling that a subframe lasts 1\;ms).

In addition to RR and MD, the performance of the following allocation algorithms, obtained by simulation, are shown. 
\begin{itemize}
	\item \textit{Centralized with resource reuse (CRR)}: applying the algorithm detailed in \cite{CecBazMasZan:C17_2}, the network allocates the resources based on the position of vehicles (which is assumed to be known with a Gaussian error of less than 100~m in 95\% of the cases). The algorithm ensures that two nodes are never assigned the same resource if their distance is less than a given value, set here at 200~m; the node is blocked (i.e., no resources are temporarily allocated) if no \ac{CAM-R} respects this condition; 
	\item \textit{Location-based graph coloring (LGC) algorithm}: applying the algorithm detailed in \cite{HuEicDilBotGoz:C16}, the network allocates the resources trying to maximize the distance between nodes using the same CAM-R; more specifically, it considers the vehicles in random order and, after allocating a different time-frequency combination to the first $\nRes$ vehicles, allocates to each of the following vehicles, one after the other, the \ac{CAM-R} that is used farther than its position, exploiting the Euclidean distance; perfect knowledge of positions and a reallocation every $\CAMperiod$ are assumed to maximize its performance;
	\item \textit{3GPP Autonomous with $\Pkeep=0$ (M4,$\Pkeep=0$) and 3GPP Autonomous with $\Pkeep=0.8$ (M4,$\Pkeep=0.8$)}: with  the 3GPP algorithm, detailed in \cite{3GPP_TS_36_213,3GPP_TS_36_321,MolGoz:J17}, the resources are allocated autonomously by the nodes, without any contribution from the network. Each node selects the resource randomly in the 20\% less interfered \acp{CAM-R},  based on the local channel sensing and limiting the choice to those that the control channel indicates as free or for which the measured interference is below a parametric threshold $I_\text{th}$ (here assumed $-110$~dB as in \cite{MolGoz:J17}); then, after a period of time randomly selected within a range of 0.5-1.5~s, the node changes allocation with probability $1-\Pkeep$, where $\Pkeep \in[0, 0.8]$ is a parameter which defines the probability of maintaining the same allocation. Since $\Pkeep$ has a significant influence on the performance \cite{TogSaiNouMugEtal:C18,BazCecZanMas:J18}, here we consider both extremes, i.e., $\Pkeep=0$ and $\Pkeep=0.8$.
\end{itemize}

All the simulations, shown together with the 95\% {\it t}-test based confidence interval, have been obtained using the LTEV2Vsim open source simulator \cite{CecBazMasZan:C17}. We also remark that in all the simulations i) the mobility is reproduced in detail, ii) all the interferers are always considered in the calculation of \ac{SINR}, iii) all the signals, including those from interferers, are affected by correlated shadowing. This allows to validate the approximations made in the analysis.

\begin{figure}[t]
	\centering
	\includegraphics[width=0.95\linewidth,draft=false]{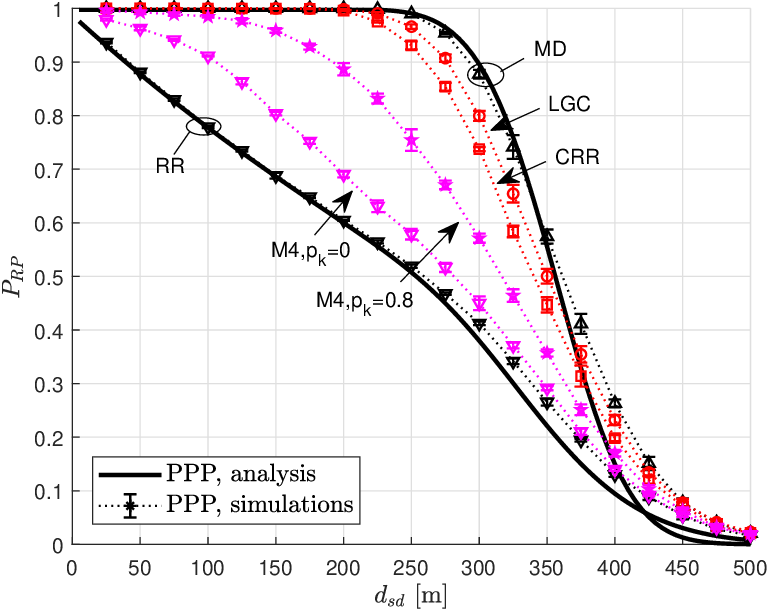}
	\caption{PPP scenario. Packet reception ratio vs. distance.}
	\label{Fig:PrrD300B}
\end{figure}

\subsection{Results in a 1-D PPP scenario}

The first set of results assumes that the vehicles are distributed following a 1-D homogeneous \ac{PPP}. In Fig.~\ref{Fig:PrrD300B}, $\prr$ is shown varying $\dist$.  Regarding RR and MD, the first observation is that the agreement between simulation and analysis is extremely good.

Focusing on \ac{MD}, Fig.~\ref{Fig:PrrD300B} presents near-threshold behaviour for $\prr$: it stays close to 1 until $\dist<300$~m and then drops rapidly to 0. 
For large values of $\dist$, other allocations (including \ac{RR}) may statistically imply the existence of a few nodes that perceive a limited amount of interference, which leads to $\prr$ better than that obtained with MD. However, this effect only occurs when the conditions of link quality are very poor and the value of $\prr$ is unacceptably small.

\begin{figure}[t]
	\centering
	\includegraphics[width=0.95\linewidth,draft=false]{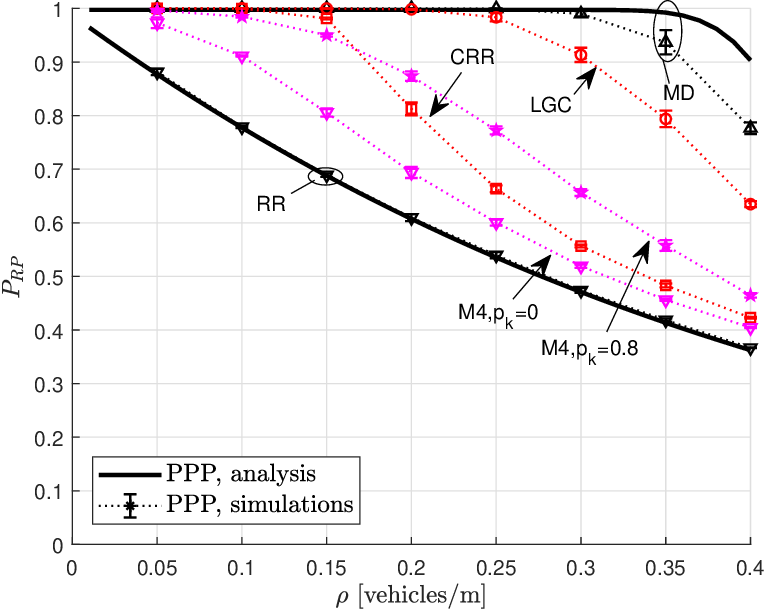}
	\caption{PPP scenario. Packet reception ratio vs. density, with  $\dist=100$~m.}
	\label{Fig:PrrRho300B}
\end{figure}

\begin{figure}[t]
	\centering
	\includegraphics[width=0.95\linewidth,draft=false]{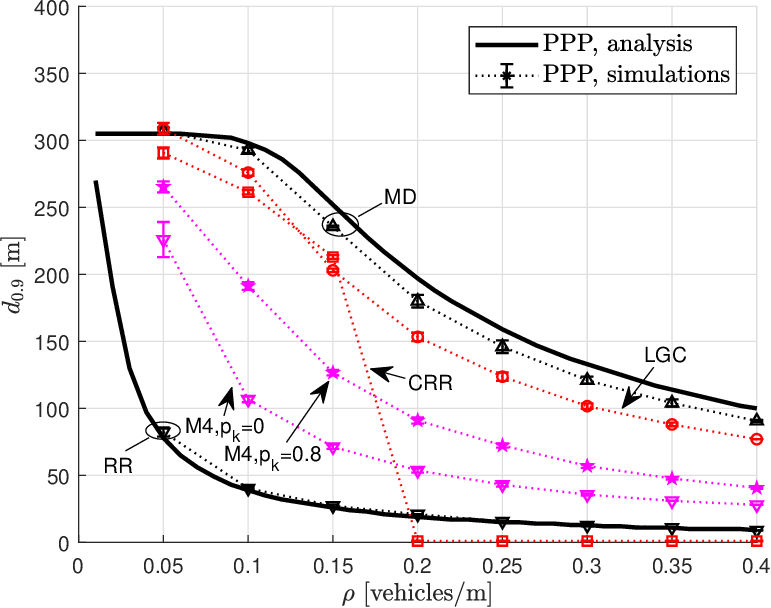}
	\caption{PPP scenario. Maximum distance allowing $\prr>0.9$ vs. density.}
	\label{Fig:MaxDrho300B}
\end{figure}

Another important aspect to highlight observing Fig.~\ref{Fig:PrrD300B} is the very high difference between \ac{MD} and \ac{RR}, with the former allowing about twice as much $\prr$ in some conditions. This confirms that the specific algorithm used for resource allocation has a significant impact on the performance of LTE-V2V. 

Focusing now on the remaining curves (i.e., CRR, LGC, M4,$\Pkeep=0$, and M4,$\Pkeep=0.8$), it is of paramount importance to observe that they all remain between RR and MD: the one providing worse performance (i.e., M4,$\Pkeep=0$) is preferable to RR and the one with the highest \ac{PRP} (i.e., LGC) does not reach MD. This confirms that the proposed algorithms can be used as references for the validation of new proposals. Less relevant to the purpose of this paper, but still interesting to note, is that the curves confirm that network controlled algorithms (i.e., CRR and LGC) in most cases lead to higher \ac{PRP} than autonomous algorithms (i.e., M4,$\Pkeep=0$ and M4,$\Pkeep=0.8$).

Similar conclusions can also be drawn from Figs.~\ref{Fig:PrrRho300B} and~\ref{Fig:MaxDrho300B}, which show, as a function of the vehicle density $\density$, respectively $\prr$ and $\distQoS$. The parameter $\distQoS$, in particular, is defined as the maximum distance between source and destination that guarantees $\prr>0.9$. As a premise, as regards CRR in Fig.~\ref{Fig:MaxDrho300B}, note that $\distQoS$ falls to zero for $\density \geq 0.2$; this can be explained by observing that the density is so high that the algorithm blocks more than 10\% of the nodes; since the blocked nodes contribute as if the corresponding messages were not received, the target $\prr>0.9$ can not be reached at any distance. The same reason, leads CRR in Fig.~\ref{Fig:PrrRho300B} to perform worse than M4,$\Pkeep=0.8$ for $\density$ greater than 0.15 vehicles/m.
Observing now both figures, the comparison between theoretical benchmarks  and simulation confirms the validity of the analysis.  In particular, it should be noted that the approximation adopted in \ac{MD} to consider only the first interferer on both sides has a negligible impact on the overall results. Moreover, also in this case we can observe that RR and MD appear as inferior and superior references for all the other algorithms.

\subsection{Results in a realistic highway scenario}

The second set of results compares the benchmark curves with simulations carried out in a realistic highway scenario. More specifically, vehicle positions are obtained from a traffic trace that represents a congested highway with three lanes per direction (details can be found in \cite{BazMasZanCal:J16}).\footnote{The trace is publicly available and can be downloaded at http://www.wcsg.ieiit.cnr.it/people/bazzi/SHINE.html.} There are on average 2015 vehicles distributed over 16~km, corresponding to a density of $\rho=0.125$~vehicles/m. The benchmark curves are obtained by applying the same density to the analysis.

\begin{figure}[t]
	\centering
	\includegraphics[width=0.95\linewidth,draft=false]{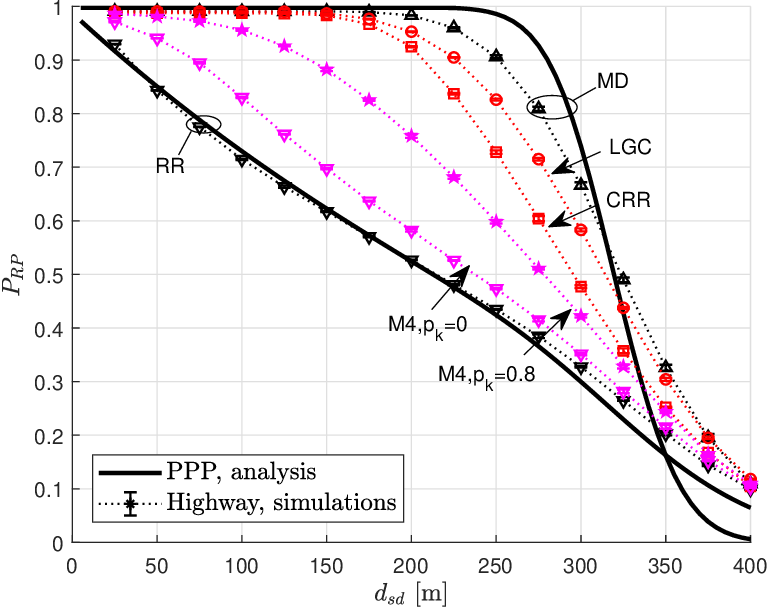}
	\caption{Highway scenario (the analysis assumes 1-D PPP with the same density). Packet reception ratio vs. distance.}\label{fig:highway}
\end{figure}

In Fig.~\ref{fig:highway}, $\prr$ is shown by varying $\dist$. The results are not generally different from those presented focusing on the 1-D PPP scenario and once again confirm that the two proposed algorithms represent valid references to evaluate the effectiveness of the resource allocation schemes. In addition to what has already been discussed, Fig.~\ref{fig:highway} allows us to derive the following consideration: although the simulation results have been obtained with a different model (realistic highway traces instead of \ac{PPP}), the theoretical reference curves are very close to those of the simulations; therefore, the curves obtained can be used not only in an ideal \ac{PPP} scenario, but also in realistic highway conditions.

\begin{figure}[t!]
	\centering
	\includegraphics[width=0.95\linewidth,draft=false]{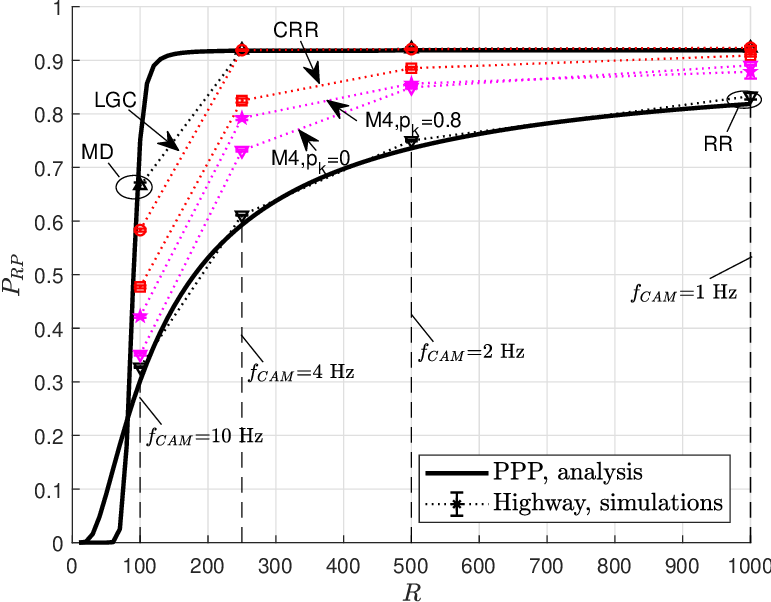}
	\caption{Highway scenario (the analysis assumes 1-D PPP with the same density). Packet reception ratio vs. number of resources that follows a variable CAM frequency, with $\dist=300$~m. }
	\label{Fig:PrrPr300B}
\end{figure}

Finally, Fig.~\ref{Fig:PrrPr300B} shows $\prr$ depending on the number of available resources $\nRes$, or equivalently, as a function of the beacon frequency, $\CAMfrequency$ (from 1 to 10 Hz). Besides confirming once again the analysis and the suitability of MD and RR as benchmarks, the figure shows that the impact of the allocation algorithm increases as resources get scarce; therefore, the design of an efficient algorithm becomes more important with an increase of $\CAMfrequency$.

\section{Conclusion and Future Work}\label{Section:conclusion}

The objective of this article is to define benchmarks that provide a clear reference to quantify the performance of the resource allocation algorithms designed for LTE-V2X. In particular, two algorithms are identified, one as an inferior and the other as a superior reference, and their performance is obtained analytically in terms of packet reception ratio in a highway scenario. The correctness of the analysis has been demonstrated with simulations, which also include aspects that are simplified in the analysis. Furthermore, the actual usability of these benchmarks has been shown using four allocations taken from the literature, which provide results that remain within the space indicated by the derived curves.
Although obtained with a 1-D PPP assumption, it has been shown that the results are valid also in realistic highway scenarios. By providing references to the performance that can be achieved, these benchmarks will be a useful tool to help in the definition and validation of new allocation algorithms.

As further improvements to the proposed analytic framework, we will work to also take into account variable beacon generation rates, given that both ETSI and SAE standards include mechanisms to make cooperative awareness message generation adaptive to channel occupation. In addition, we will extend the investigation to other relevant metrics, such as latency and update delay (distribution of the delay between consecutive updates of a given vehicle).

\section*{Appendix A: Proof of Proposition 1}
Let us consider the set of nodes, characterized by the positions $x_k$, which adopt the same resource $r$. For the considerations expressed in Section \ref{Subsection:randomcalculation}, this set still belongs to a 1-D homogeneous \ac{PPP} with density $\densityRR$. Now, let $\PIi{k}$ denote the interference received by the generic node $k$, at position $x_k$, provided by a transmitter at position $x=0$. Defining
\begin{equation}\label{eq:Pr0}
\PRZERO \triangleq \frac{\PTX \Gt \Gr}{\alphaPar{}}
\end{equation}
the received power at 1~m, it follows that the interference is equal to
\begin{equation}\label{eq:G}
\PIi{k} = \G{|x_k|} \triangleq \PRZERO |x_k|^{-\beta}\;
\end{equation}
and the interference provided by all the nodes that use the same resource $r$ can be written as 
\begin{align}\label{eq:I1}
\PI = \sum_{k=-\infty}^{\infty} \PIi{i}= \sum_{k=-\infty}^{\infty} \G{|x_k|} \;.
\end{align}

To obtain the \ac{CDF} of $\PI$, we will first evaluate its \ac{c.f.}. 
To this aim, let us consider a  finite interval $[-a,a]$. First we provide the \ac{c.f.} for the finite interval and, then, calculate the limit for $a\rightarrow\infty$. Since nodes are distributed according to a homogeneous \ac{PPP}, if $\PhiY(\omega)$ is the \ac{c.f.} of $\PI$, then \cite{SalZanJ09}
\begin{equation}
\label{eq:phiya1}
\PhiYa(\omega)=e^{\mu_N (\PhiGa(\omega)-1)}
\end{equation}
where $\PhiGa(\omega)$ is the \ac{c.f.} of interference generated by a generic node in $[-a,a]$ and $\mu_N=2 a \densityRR$ is the expected value of the number of nodes in the interval.

To calculate $\PhiGa(\omega)$, recall that the \ac{PDF} of the distance between a generic interfering node and the receiver located at $x=0$ is
\begin{equation}
f_{\text{R}_a}(\vardist)=1/a  ,\,\,\,\vardist \in [0,a].
\end{equation}
In such case,  it is 
\begin{align}
\PhiGa(\omega)
&=\EXs{{\text{R}}_a}{e^{j \omega \G{|x|}}}=\int_{0}^{a}f_{\text{R}_a}(\vardist) e^{j \omega \G{\vardist}}d\vardist
\nonumber\\&
=\frac{1}{a}\int_{0}^{a} e^{j \omega \G{\vardist}}d\vardist
\end{align}
which becomes, after the change of variable $\varInterf=\G{\vardist}$, $\vardist=\Ginv{\varInterf}$, $d \vardist=[\Ginv{\varInterf}]' dx$ and an integration by parts
\begin{align}
\label{eq:phiga1}
\PhiGa(\omega)&=\frac{1}{a}\int_{\G{0}}^{\G{a}} e^{j \omega \varInterf} [\Ginv{\varInterf}]' d\varInterf
\nonumber\\&
= \frac{1}{a} \left[ \left[ \Ginv{\varInterf} e^{j \omega \varInterf}\right]_{\G{0}}^{\G{a}}  - j \omega \int_{\G{0}}^{\G{a}} \Ginv{\varInterf} e^{j \omega \varInterf} d\varInterf   \right]\nonumber\\&= \frac{1}{a} \left[  a e^{j \omega \G{a}} - j \omega \int_{\G{0}}^{\G{a}} \Ginv{\varInterf} e^{j \omega \varInterf} d\varInterf   \right]\;.
\end{align}

By substituting \eqref{eq:phiga1} in \eqref{eq:phiya1}, we get
\begin{align}
\PhiYa(\omega)&=e^{\mu_N (\PhiGa(\omega)-1)}
\nonumber\\&
=e^{2 \densityRR a \left[ \frac{1}{a} \left( a e^{j \omega \G{a}} - j \omega \int_{\G{0}}^{\G{a}}  \Ginv{\varInterf}  e^{j \omega \varInterf}d\varInterf   \right) -1 \right]  }\nonumber\\&=e^{2 \densityRR a \left[  \left( e^{j \omega \G{a}}-1 \right) - \frac{j \omega}{a} \int_{\G{0}}^{\G{a}}  \Ginv{\varInterf} e^{j \omega \varInterf} d\varInterf    \right]  }
\nonumber\\&
=e^{2 \densityRR a \left( e^{j \omega \G{a}}-1 \right)} e^{-j 2 \densityRR  \omega \int_{\G{0}}^{\G{a}}  \Ginv{\varInterf} e^{j \omega \varInterf} d\varInterf}.
\end{align}

If we now apply the limit for $a \rightarrow \infty$, we obtain
\begin{align}
\PhiY(\omega)&=\lim\limits_{a\rightarrow \infty}\PhiYa(\omega)
\nonumber\\&
=\lim\limits_{a\rightarrow \infty} e^{2 \densityRR a \left( e^{j \omega \G{a}}-1 \right)} e^{-j 2 \densityRR  \omega \int_{\G{0}}^{\G{a}}  \Ginv{\varInterf} e^{j \omega \varInterf} d\varInterf}
\end{align}
which, recalling the definition of $\G{r}$ in  \eqref{eq:G}, simplifies to
\begin{align}
\PhiY(\omega)&=e^{-j 2 \densityRR  \omega \int_{\G{0}}^{\G{\infty}}  \Ginv{\varInterf} e^{j \omega \varInterf} d\varInterf}\;.
\end{align}

Equation \eqref{eq:G} also implies that $\Ginv{\varInterf} = {\frac{\varInterf}{\PRZERO}}^{-1/\betaPar{}}$, $\G{0}\rightarrow \infty$, and $\G{\infty}=0$; thus we can write
\begin{align}\label{eq:PhiY2}
\PhiY(\omega)&=
e^{-j 2 \densityRR  \omega \int_{\infty}^{0}  {\frac{\varInterf}{\PRZERO}}^{-1/\betaPar{}} e^{j \omega \varInterf} d\varInterf}\nonumber\\&=
e^{j 2 \densityRR  \omega \int_{0}^{\infty}  {\frac{\varInterf}{\PRZERO}}^{-1/\betaPar{}} e^{j \omega \varInterf} d\varInterf}
\nonumber\\&=e^{j 2 \densityRR  \omega \int_{0}^{\infty}  {\frac{\varInterf}{\PRZERO}}^{-1/\betaPar{}} \cos{\left(\omega \varInterf\right)} d\varInterf} \nonumber\\&\;\;\;\;\;\cdot e^{- 2 \densityRR  \omega \int_{0}^{\infty}  {\frac{\varInterf}{\PRZERO}}^{-1/\betaPar{}}\sin{\left(\omega \varInterf\right)} d\varInterf }\nonumber\\&=
e^{-2\densityRR \PRZERO^{1/\betaPar{}} \cos\left(\frac{\pi}{2 \beta}\right)\Gamma\left(\frac{\betaPar{}-1}{\betaPar{}}\right)\omega|\omega|^\frac{1-\betaPar{}}{\betaPar{}}\text{sign}(\omega)}
\nonumber\\&\;\;\;\;\;\cdot 
e^{j2\densityRR \PRZERO^{1/\betaPar{}} \sin\left(\frac{\pi}{2 \betaPar{}}\right)\Gamma\left(\frac{\betaPar{}-1}{\betaPar{}}\right)\omega|\omega|^\frac{1-\betaPar{}}{\betaPar{}}} 
\;.
\end{align}

Defining $$\KRR\triangleq 2\densityRR \PRZERO^{1/\betaPar} \Gamma\left(\frac{\betaPar-1}{\betaPar}\right)$$ we then obtain
\begin{align}\label{eq:PhiY3}
\PhiY(\omega)=
e^{-\KRR |\omega|^\frac{1}{\betaPar} \left[\cos\left(\frac{\pi}{2 \betaPar}\right)-j \sin\left(\frac{\pi}{2 \betaPar}\text{sign}(\omega)\right)\right]}
\end{align}
which can be rearranged as 
\begin{align}\label{eq:PhiY4}
\PhiY(\omega)&=
e^{-\KRR \cos\left(\frac{\pi}{2 \betaPar}\right) |\omega|^\frac{1}{\betaPar} \left[1-j \frac{\sin\left(\frac{\pi}{2 \betaPar}\right)}{\cos\left(\frac{\pi}{2 \betaPar}\right)}\text{sign}(\omega)\right]}
\nonumber\\&
= e^{-|\left(k \cos\left(\frac{\pi}{2 \betaPar}\right) \right)^\betaPar \cdot \omega|^\frac{1}{\betaPar} \left[1-j \cdot 1 \cdot \text{sign}(\omega) \cdot \tan\left(\frac{\pi}{2 \betaPar}\right)\right]}
\;.
\end{align}

Equation \eqref{eq:PhiY4} corresponds to the characteristic function of a stable distribution 
\begin{align}\label{eq:stable}
\varphi(t;a,b,c,\mu)=e^{j t \mu - |c t|^a \left(1-j b\; \text{sign}(t) \tan\left(\frac{\pi a}{2}\right)\right))}
\end{align}
with variable $t=\omega$ and parameters $a=1/\betaPar$, $b=1$, $c=\left(\KRR \cos\left(\frac{\pi}{2 \betaPar}\right) \right)^\betaPar$, and $\mu=0$. This demonstrates Proposition~1. 

\section*{Appendix B: Proof of Lemma 1}
	In the case of $\betaPar=2$, the \ac{c.f.} in \eqref{eq:PhiY4} becomes
	\begin{align}\label{eq:PhiY_beta2}
	\PhiY(\omega)=
	e^{ |\frac{\KRR^2}{2} \omega|^\frac{1}{2} \left(1-j \text{sign}(\omega)\right)}
	\end{align}
	which corresponds to the characteristic function of a Levy distribution, with location parameter $\mu=0$ and scale parameter $c=\frac{\KRR^2}{2}=2 \PRZERO \left(\densityRR  \Gamma\left(1/2\right)\right)^2$. This leads to \eqref{eq:Flemma}, thus demonstrating Lemma 1.

\section*{Appendix C: Proof of Proposition 2}
Let us indicate with $\mathcal{N}$ the set of all vehicles, with $d_{ij}$ the distance between the generic nodes $i \in \mathcal{N}$ and $j \in \mathcal{N}$, with $\res{i}$ the resource allocated to $i$, and with $\mathcal{N}_i$ the set of nodes in $\mathcal{N}-\left\{i\right\}$ that uses the same resource as $i$. Proposition 2 states that
the average distance between nodes using the same resource, defined as 
\begin{align}
\distaverage = \frac{ \sum_{i\in \mathcal{N}}\sum_{j \in \mathcal{N}_i} d_{ij}}{ \sum_{i\in \mathcal{N}} |\mathcal{N}_i| }
\end{align}
where $|\mathcal{A}|$ is the cardinality of the set $\mathcal{A}$, is maximized with MD. The demonstration is provided by contradiction and shows that, starting with MD, any change reduces $\distaverage$. Let us focus on the generic node $a$, which is allocated to resource $\res{A} = (a\mod\nRes)$. With MD, the first node on the right of $a$ that uses the same resource as $a$ is node $b = a+\nRes$. If the allocation of user $a$ is modified to any resource in $\{1,2,...,\nRes\}-\{\res{A}\}$, there is exactly one node  $c$ such that $x_a<x_c<x_b$. Since $x_c-x_a < x_b-x_a$, the contribution to the average distance involving the first node from $a$ on the right has decreased. Since the same reasoning leads to the same conclusion for each additional node that uses the resource $\res{A}$ on the right of $a$ and for each node that uses the resource $\res{A}$ on the left of $a$, it is proved that any variation in the allocation of an arbitrary node reduces the average distance between the nodes using the same resource.

\section*{Appendix D: Proof of Proposition 3}
\subsection{Introduction} Given a generic receiver at distance $\dist > 0$ from a source $i$ in the position $x_i$, the total interference to the receiver with MD is
\begin{align}\label{eq:IMD1}
\PI = \sum_{j=-\infty,j\neq i}^{\infty} \G{|x_{i + j \nRes}-x_i-\dist|}\;.
\end{align}
We limit our attention to situations where the receiver is to the right of the transmitter, thus in position $x_i+\dist$. Please observe that, given the symmetry of the scenario, the case with the receiver on the left side occurs with probability 0.5 and leads to the same probabilities, so the conclusion remains valid even removing this hypothesis.

\subsection{Statistic of the positions} As a first step, it is useful to calculate the \ac{PDF} and \ac{CDF} of the distance of the n-th node from the source. Given the assumption of 1-D \ac{PPP} distribution of nodes, the \ac{PDF} of the distance between a node and its \mbox{$n$-th} neighbor in a specific direction (i.e., left or right) can be calculated with the same approach as \cite{Hae:J05}. Denoting $\vardistn$ the \ac{r.v.} of the $n$-th node distance, its \ac{CCDF} can be calculated as 
\begin{align}\label{eq:ccdfdistn}
\Fcompldistn(\delta) &= \Prob\{\text{less than}~n~\text{nodes} \in \left[0,\delta\right]\} 
\nonumber \\&
=\sum_{k=0}^{n-1} \frac{\left(\density\vardist\right)^k}{k!} e^{-\density \vardist} =\frac{\Gamma(n,\rho \vardist)}{\Gamma(n)}
\end{align}
and the corresponding \ac{CDF} is
\begin{align}\label{eq:Fr}
\Fdistn(\vardist) = 1 - \frac{\Gamma(n,\rho \vardist)}{\Gamma(n)}\;.
\end{align}

From \eqref{eq:ccdfdistn} the \ac{PDF} can be obtained as\footnote{The same result can be obtained from  \cite[eq. (21)]{ZanBazPasMas:J13}, assuming $j=n$, $\mu/a=\rho$ and $a \rightarrow \infty$.} 
\begin{align}\label{eq:pdfdistn}
\fdistn(\delta) &= -\frac{d \Fcompldistn(\delta)}{d \delta} 
\nonumber \\&
=-\left[-\density e^{-\density \vardist} \sum_{k=0}^{n-1} \frac{\left(\density\vardist\right)^k}{k!} + e^{-\density \vardist} \sum_{k=1}^{n-1} \frac{k\left(\density\right)^k\left(\vardist\right)^{k-1}}{k!}  \right]\nonumber \\
&=\density e^{-\density \vardist}\left[\sum_{k=0}^{n-1} \frac{\left(\density\vardist\right)^k}{k!} + \sum_{k=0}^{n-2} \frac{\left(\density \vardist\right)^{k}}{k!}  \right]=\density e^{-\density \vardist} \frac{\left(\density\vardist\right)^{n-1}}{(n-1)!}
\nonumber \\
&=\frac{\density^n}{\Gamma(n)} \vardist^{n-1} e^{-\density \vardist}\;. 
\end{align}
Equations \eqref{eq:ccdfdistn}, \eqref{eq:Fr} and \eqref{eq:pdfdistn} are valid for any $\vardist\geq 0$.

\begin{table}[!t]
	\caption{Values of $\paramA$ and $\paramB$.}
		\label{Tab:paramsAB}
	\centering
	\begin{tabular}{p{4.5cm}|p{0.3cm}p{0.3cm}p{2cm}}
	\textbf{Conditions} & \textbf{$\paramA$} & \textbf{$\paramB$} & \textbf{$\setValidy$} \\ \hline 
	Interferer at the left of the source & +1 & +1 & 
	$\left(0,\G{\dist}\right)$ \\
	Interferer between source and receiver & -1 & +1 & $\left(\G{\dist},\infty\right)$ \\
	Interferer at the right of the receiver & +1 & -1 & $\left(0,\infty\right)$\\
	\hline
	\end{tabular}
\end{table}

\subsection{Statistic of the interference from a generic node} The second step consists in calculating the distribution of the interference generated by a node in the position $x_j$ with respect to a useful signal generated by the position $x_i$ and a receiver located in  $x_i+\dist$. For this purpose, let us first define the variable $\paramA = \text{sign}(x_i-x_j)\cdot\text{sign}\left(d-(x_j-x_i)\right)$ (where $\text{sign}(x)$ is the sign of $x$), which indicates if the interferer is between the source and receiver ($\paramA=-1$) or not ($\paramA=1$), and $\paramB = \text{sign}\left(d-(x_j-x_i)\right)$, which indicates if the interferer is to the left of the source ($\paramB=+1$) or to its right ($\paramB=-1$).

Recalling \eqref{eq:G} and using the definitions of $\paramA$ and $\paramB$, the interference as a function of $\vardist=|x_j-x_i|$ can be calculated for any $\vardist>0$ as
\begin{equation}\label{eq:gshifted}
\GshiftedAlfa{\vardist} = \G{\paramA \vardist + \paramB \dist} = \PRZERO {\left(\paramA \vardist + \paramB \dist\right)}^{-\betaPar}
\end{equation}
and its inverse results in
\begin{equation}\label{eq:gshiftedinv}
\GshiftedAlfainv{\varInterf} = \paramA \left[{\left(\frac{\varInterf}{\PRZERO}\right)}^{-1/\betaPar}-\paramB  \dist\right]\;.
\end{equation}
Additionally, the derivative of the inverse is
\begin{equation}\label{eq:gshiftedinvder}
[\GshiftedAlfainv{\varInterf}]' = -\frac{\paramA}{\betaPar}\PRZERO^{1/\betaPar}\varInterf^{-\frac{1}{\betaPar}-1}\;.
\end{equation}
Both \eqref{eq:gshiftedinv} and \eqref{eq:gshiftedinvder} are valid for any positive $y$ so that $\paramA \left[{\left(\frac{\varInterf}{\PRZERO}\right)}^{-1/\betaPar}-\paramB  \dist\right]>0$. Using $\setValidy$ to describe the set in which $y$ is valid, it results: 1) $\setValidy=\left(0,\G{\dist}\right)$ when $\paramA=+1$ and $\paramB=+1$; 2) $\setValidy=\left(\G{\dist},\infty\right)$ when $\paramA=-1$ and $\paramB=+1$; and 3) $\setValidy=\left(0,\infty\right)$ when $\paramA=+1$ and $\paramB=-1$. All possible combinations of  $\paramA$ and $\paramB$, and the resulting $\setValidy$ are summarized in Table~\ref{Tab:paramsAB}.

Using \eqref{eq:pdfdistn}, \eqref{eq:gshiftedinv}, \eqref{eq:gshiftedinvder} and the change-of-variable technique, the \ac{PDF} of the interference from the $n$-th node can be calculated as in the following equation, where the dependence on $\paramA$ and $\paramB$ is explicit.
\begin{align}\label{eq:pdfIlpartialgeneric}
f_{\PIgeneric}(\varInterf) &= -\fdistn(\GshiftedAlfainv{\varInterf}) \cdot [\GshiftedAlfainv{\varInterf}]'
\nonumber\\&
= -\frac{\density^n}{\Gamma(n)} \left(\GshiftedAlfainv{\varInterf}\right)^{n-1} e^{-\density \GshiftedAlfainv{\varInterf}} [\GshiftedAlfainv{\varInterf}]' \nonumber\\&=-\frac{\density^n}{\Gamma(n)} \left(\paramA\left(\frac{\varInterf}{\PRZERO}\right)^{-1/\betaPar}-\paramB\dist\right)^{n-1}
\nonumber\\&\;\;\;\;\;\;\;\;\;\cdot
e^{-\density \left(\paramA\left(\frac{\varInterf}{\PRZERO}\right)^{-1/\betaPar}-\paramB\dist\right)}\left(-\frac{\paramA}{\betaPar}\right)\PRZERO^{1/\betaPar}\varInterf^{-\frac{\betaPar+1}{\betaPar}}\;.
\end{align}


Thus, defining $$\subfuncMDalpha{a}{b}\triangleq\paramA\left(\frac{a}{\PRZERO}\right)^{-\frac{1}{\betaPar}}-\paramB b$$ and $$\subfuncMDalphader{a} \triangleq \frac{d\subfuncMDalpha{a}{b}}{d a}=\left(-\frac{\paramA}{\betaPar}\right)\PRZERO^{1/\betaPar}a^{-\frac{1+\betaPar}{\betaPar}}$$ 
we get
\begin{align}\label{eq:pdfIlgeneric}
f_{\PIgeneric}(y) = \left\{ \begin{array}{ll}  
-\frac{\density^n}{\Gamma(n)} 
\left(\subfuncMDalpha{\varInterf}{\dist}\right)^{n-1} & \\ \;\;\;\;\;\;\;\;\cdot
\subfuncMDalphader{\varInterf}e^{-\density \left(\subfuncMDalpha{y}{\dist}\right)}  & \varInterf \in \setValidy \\ 0 & \text{otherwise}
\end{array} \right.\;.
\end{align}
From \eqref{eq:pdfIlgeneric}, we can also write the corresponding \ac{CDF} as
\begin{align}\label{eq:cdfIlgeneric}
F_{\PIgeneric}(\varInterf) &= \int_{0}^\varInterf f_{\PIgeneric}(z) dz \nonumber\\&
= \left\{ \begin{array}{ll}  \frac{\Gamma\left(n,\density \left(\subfuncMDalpha{\varInterf}{\dist}\right)\right)}{\Gamma(R)} & \varInterf \in \setValidy \\ 0 & \varInterf < \min\{\setValidy\} \\ 1 & \varInterf > \max\{\setValidy\}
\end{array} \right.\;.
\end{align}

\begin{figure} [t]
	\centering
	\subfigure[Example of case 1.]{
		\includegraphics[trim={200 620 200 0}, clip,width=0.8\linewidth,draft=false]{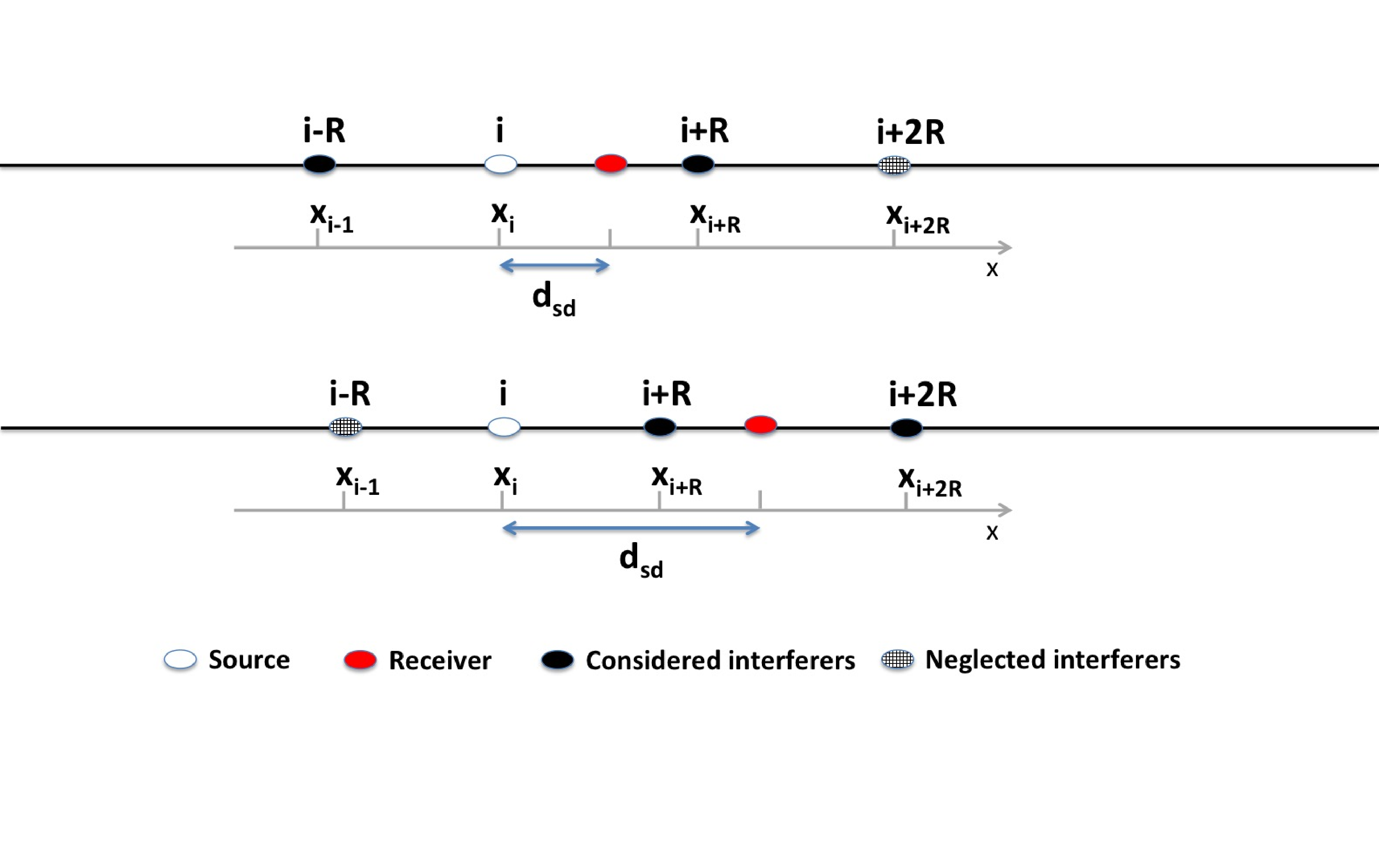}\label{fig:MD_a}}
	\subfigure[Example of case 2.]{
		\includegraphics[trim= {200 300 200 400}, clip,width=0.8\linewidth,draft=false]{Figures/Diapositiva2.eps}\label{fig:MD_b}}
	\subfigure[Legend.]{
		\includegraphics[trim= {150 200 150 730}, clip,width=0.95\linewidth,draft=false]{Figures/Diapositiva2.eps}\label{fig:MD_c}}
	\caption{MD: examples of case 1 and case 2.}
	\label{fig:MD}
\end{figure}

\subsection{Statistic of the overall interference} Given these premises, we can now proceed to calculate the distribution of the overall interference. For this purpose, please recall that all potential interferers are the $k \cdot R$-th farthest nodes to the right and left of the source, with $k$ being any positive integer. If we neglect the interference from all the nodes except the interferers closest to the receiver from both directions, hereafter called left and right interferers, we can rewrite \eqref{eq:IMD1} as
\begin{align}\label{eq:I2}
\PI \simeq \G{{\vardist}^{(l)}} + \G{{\vardist}^{(r)}}\;
\end{align}
where the values of ${\vardist}^{(l)}$ and ${\vardist}^{(r)}$ depend on the position of the receiver with respect to the source and to the interferers. In particular, let us neglect the case in which $\dist \geq x_{i + 2 \nRes}-x_{i}$\footnote{The case in which $\dist \geq x_{i + 2 \nRes}-x_{i}$ occurs with probability $\Fdist{2\nRes}(\dist)$; from \eqref{eq:ccdfdistn}, if we assume for example $\nRes=100$, $\density=0.1$, and $\dist=500$~m, it is less than $10^{-9}$.} and separate the two following cases, exemplified in Fig.~\ref{fig:MD}:
\begin{itemize}
	\item Case 1: $\dist < x_{i + \nRes}-x_{i}$, which means that the receiver is between the source and the first interferer to the left of the source (Fig.~\ref{fig:MD_a}); this case occurs with probability $\cFdistR(\dist)$ (given in \eqref{eq:ccdfdistn});
	\item Case 2: $x_{i + \nRes}-x_{i} \leq \dist < x_{i + 2 \nRes}-x_{i}$, which means that the receiver is between the first interferer to the right and the second interferer to the right of the source (Fig.~\ref{fig:MD_b}); this case occurs with probability $\FdistR(\dist)$.
\end{itemize}
In the following, we first obtain the distribution of the left and right interference and then derive the \ac{CDF} of $\PI$.

Using \eqref{eq:pdfIlgeneric} and \eqref{eq:cdfIlgeneric}, the interference caused by the first interferer to the left and to the right of the receiver for both Case 1 and Case 2 can be calculated as follows.

\textit{Left interferer.} In Case 1, the interferer to the left of the receiver is node $i-R$, whereas in Case 2 it is node $i+R$. Thus, applying the law of total probability and using \eqref{eq:pdfIlgeneric}, it is
\begin{align}\label{eq:pdfPIleft}
f_{\PIleft}(\varInterf) &= 
\cFdistR(\dist) f_{\PIgenericTriple{R,+1,+1}}(\varInterf) + \FdistR(\dist) f_{\PIgenericTriple{R,-1,+1}}(\varInterf)
\end{align}
and the corresponding \ac{CDF} is
\begin{align}\label{eq:cdfPIleft}
F_{\PIleft}(\varInterf) &= 
	\cFdistR(\dist) F_{\PIgenericTriple{R,+1,+1}}(\varInterf) + \FdistR(\dist) F_{\PIgenericTriple{R,-1,+1}}(\varInterf)\;.
\end{align}

\textit{Right interferer.} Similarly, the interferer to the left of the receiver is node $i+R$, whereas in Case 2 it is node $i+2R$, thus it is
\begin{align}\label{eq:pdfPIright}
f_{\PIright}(\varInterf) &=
	\cFdistR(\dist) f_{\PIgenericTriple{R,+1,-1}}(\varInterf) + \FdistR(\dist) f_{\PIgenericTriple{2R,+1,-1}}(\varInterf)\;.
\end{align}

\textit{Statistic of the sum.} Finally, given the \ac{CDF} of the interference $F_{\PIleft}(\varInterf)$ received from the first interferer to the left in \eqref{eq:cdfPIleft} and 
the \ac{PDF} of the interference $f_{\PIright}(\varInterf)$ received from the first interferer to the right in \eqref{eq:pdfPIright}, and assuming that the two distributions are independent of each other,\footnote{This assumption is strictly true for Case~1 and an approximation for Case~2.} the \ac{CDF} of the sum interference can be written as 
\footnotesize
\begin{align}\label{eq:cdfIMDderivedComplete}
F_{\PI}(\varInterf) &= \int_{-\infty}^\infty F_{\PIleft}(\varInterf-z) f_{\PIright}(z) dz 
\nonumber\\& 
=\int_{0}^y
	\left[\cFdistR(\dist) F_{\PIgenericTriple{R,+1,+1}}(\varInterf-z) + \FdistR(\dist) F_{\PIgenericTriple{R,-1,+1}}(\varInterf-z)\right]\cdot
\nonumber\\&\left[ \cFdistR(\dist) f_{\PIgenericTriple{R,+1,-1}}(z) + \FdistR(\dist) f_{\PIgenericTriple{2R,+1,-1}}(z)
\right] dz
\nonumber\\& 
=\left(\cFdistR(\dist)\right)^2 \int_{0}^y
	 F_{\PIgenericTriple{R,+1,+1}}(\varInterf-z) f_{\PIgenericTriple{R,+1,-1}}(z) dz
\nonumber\\& 
+ \cFdistR(\dist) \FdistR(\dist) \int_{0}^y
	F_{\PIgenericTriple{R,+1,+1}}(\varInterf-z)  f_{\PIgenericTriple{2R,+1,-1}}(z) dz
\nonumber\\& 
+ \cFdistR(\dist) \FdistR(\dist) \int_{0}^y
	 F_{\PIgenericTriple{R,-1,+1}}(\varInterf-z) f_{\PIgenericTriple{R,+1,-1}}(z) dz
\nonumber\\& 
+ \left(\FdistR(\dist)\right)^2 \int_{0}^y
	F_{\PIgenericTriple{R,-1,+1}}(\varInterf-z) f_{\PIgenericTriple{2R,+1,-1}}(z) dz
\end{align}
\normalsize

In most cases of interest, it is $\FdistR(\dist) << 1$ (i.e., the probability that there is an interferer between source and destination is negligible), which also implies $\cFdistR(\dist) \simeq 1$; under this hypothesis and recalling \eqref{eq:cdfIlgeneric}, it follows
\begin{align}\label{eq:cdfIMDderivedApprox1}
F_{\PI}(\varInterf) &\approx\left(\cFdistR(\dist)\right)^2 \int_{0}^y F_{\PIgenericTriple{R,+1,+1}}(\varInterf-z) f_{\PIgenericTriple{R,+1,-1}}(z) dz \nonumber \\
&=
	1 \cdot \left[
	\int_{0}^{\min\left(y,\G{\dist}\right)}\frac{\Gamma\left(\nRes,\density\left(\subfuncMDalphaTriple{R,+1,+1}{\varInterf-z}{\dist}\right)\right)}{\Gamma(R)}\right.\nonumber\\
&\left.
	 \cdot f_{\PIgenericTriple{R,+1,-1}}(z) dz
	+
	\int_{\min\left(y,\G{\dist}\right)}^{y}1\cdot f_{\PIgenericTriple{R,+1,-1}}(z) dz
 \right]\;.
\end{align}
Furthermore, in practical cases it is normally $\varInterf<\G{\dist}$, i.e., the useful received power is greater than the minimum acceptable interference. Thus, exploiting \eqref{eq:pdfIlgeneric}, we can write
\begin{align}\label{eq:cdfIMDderivedApprox2}
F_{\PI}(\varInterf) &\approx
	\int_{0}^y \frac{\Gamma\left(\nRes,\density \left(\subfuncMDalphaTriple{R,+1,+1}{\varInterf-z}{\dist}\right)\right)}{\Gamma(R)}\nonumber\\
&\cdot \left(-\frac{\density^\nRes}{\Gamma(\nRes)}\right) 
	\left(\subfuncMDalphaTriple{R,+1,-1}{z}{\dist}\right)^{\nRes-1}\nonumber\\
&\cdot \subfuncMDalphaderTriple{R,+1}{z} e^{-\density \left(\subfuncMDalphaTriple{R,+1,-1}{z}{\dist}\right)} dz\;.
\end{align}

Finally, defining $\KMD \triangleq -\frac{\density^\nRes}{(\Gamma(\nRes))^2}$, $\subfuncMD{a}{b} \triangleq \subfuncMDalphaTriple{R,+1,-1}{a}{b}$ (which also implies $\subfuncMD{a}{b} = \subfuncMDalphaTriple{R,+1,+1}{a}{-b}$), and $\subfuncMDder{a} \triangleq \subfuncMDalphaderTriple{R,+1}{a}$, \eqref{eq:cdfIMDderivedApprox2} can be rearranged as in \eqref{eq:cdfIMD}.

Please remark that \eqref{eq:cdfIMD} holds under the assumptions that the probability that there is an interferer between source and destination is negligible and that the useful received power is greater than the minimum acceptable interference. In the rare case that they are not valid, the solution is provided by \eqref{eq:cdfIMDderivedComplete}.

\bibliographystyle{IEEEtran}
\bibliography{biblioSelf,biblioOthers,biblioStandards,connectivity}

\begin{IEEEbiography}[{\includegraphics[width=1in,height=1.25in,clip,keepaspectratio]{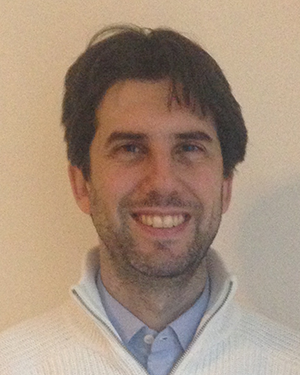}}]{Alessandro Bazzi} (S'03-M'06-SM'18) received the Laurea degree and the Ph.D. degree in telecommunications engineering both from the University of Bologna, Italy, in 2002 and 2006, respectively. Since 2002, he works with the Institute of Electronics, Computer and Telecommunication Engineering (IEIIT) of the National Research Council of Italy (CNR) and since the academic year 2006/2007, he has been acting as adjunct Professor at the University of Bologna. His work mainly focuses on connected vehicles and heterogeneous wireless access networks, with particular emphasis on medium access control, routing and radio resource management. Dr. Bazzi serves as a Reviewer and TPC Member for various international journals and conferences and he is currently in the Editorial Board of Hindawi's Mobile Information Systems and Wireless Communications and Mobile Computing.
\end{IEEEbiography}

\begin{IEEEbiography}[{\includegraphics[width=1in,height=1.25in,clip,keepaspectratio]{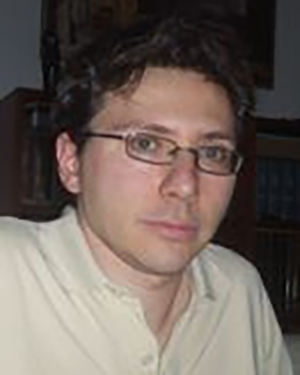}}]{Alberto Zanella} (S'99-M'00-SM'12) received the Laurea degree (summa cum laude) in Elecronic Engineering from the University of Ferrara, Italy, in 1996, and the Ph.D. degree in Electronic Engineering and Computer Science from the University of Bologna in 2000. In 2001 he joined the CNR-CSITE (merged in CNR-IEIIT since 2002) as a researcher and, since 2006, as senior researcher. His research interests include MIMO, mobile radio systems, ad hoc and sensor networks, vehicular networks. Since 2001 he has the appointment of Adjunt Professor of Electrical Communication (2001 - 2005), Telecommunication Systems (2002-2012-2013), Multimedia Communication Systems (2006 - 2011) at the University of Bologna. He participated/participate to several national and European projects. He was Technical Co-Chair of the PHY track of the IEEE conference WCNC 2009 and of the Wireless Communications Symposium (WCS) of IEEE Globecom 2009. He was/is in the Technical Program Committee of several internation conferences, such as ICC, Globecom, WCNC, PIMRC, VTC. He had served as Editor for Wireless Systems (2003-2012), IEEE Transactions on Communications and he is currently Senior Editor for the same journal.
\end{IEEEbiography}

\begin{IEEEbiography}[{\includegraphics[width=1in,height=1.25in,clip,keepaspectratio]{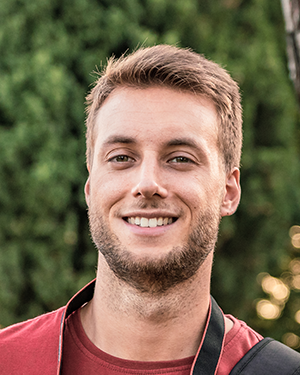}}]{Giammarco Cecchini} (S'17) received the Laurea degree (summa cum laude) in Telecommunications Engineering in Bologna in 2016, awarded with a prize from Lions Club Fondazione Guglielmo Marconi for his master thesis entitled "Performance of LTE-V2V for Cooperative Awareness". Since 2017, he joined the Institute of Electronics, Computer and Telecommunication Engineering (IEIIT) of the National Research Council of Italy (CNR). His work mainly focuses on wireless technologies for connected vehicles, both including IEEE 802.11p (with related standards) and LTE-V2X. He is also the main developer of the open-source MATLAB simulator LTEV2Vsim.
\end{IEEEbiography}

\begin{IEEEbiography}[{\includegraphics[width=1in,height=1.25in,clip,keepaspectratio]{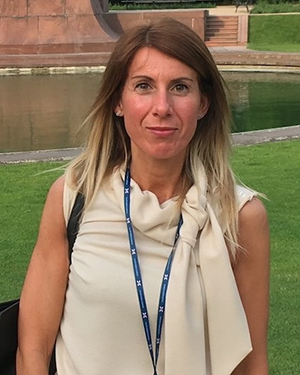}}]{Barbara M. Masini} (S'02-M'05) received the Laurea degree (summa cum laude) in Telecommunications Engineering and the Ph.D. degree in Electronic, Computer Science, and Telecommunication engineering from the University of Bologna, Italy, in 2001 and 2005, respectively. Since 2005, she is a researcher at the Institute for Electronics and for Information and Telecommunications Engineering (IEIIT), of the National Research Council (CNR). Since 2006 she is also adjunct Professor at the University of Bologna. She works in the area of wireless communication systems and her research interests are mainly focused on connected vehicles, from physical and MAC levels aspects up to applications and field trial implementations. Research is also focused on relay assisted communications, energy harvesting, and visible light communication (VLC). She is Editor of Elsevier Computer Communication, Guest Editor of Elsevier Ad Hoc Networks, Special Issue on Vehicular Networks for Mobile Crowd Sensing (2015), Mobile Information Systems, Special Issue on Connected Vehicles: Applications and Communication Challenges (2017), Sensors, Special Issue on Sensors Networks for Smart Roads (2018). She is Secretary of Chapter VT06/COM19 of the IEEE Italy section. She is TPC member of several conferences, reviewer for most international journals and for the Italian Ministry of Economic Development (MISE).
\end{IEEEbiography}
\end{document}